\documentclass[aps,pra,twocolumn,superscriptaddress]{revtex4-2}
\usepackage{graphicx}
\usepackage{amssymb,amsmath,amsthm}
\usepackage{ulem}
\usepackage{bm,bbm}
\usepackage[bookmarks=false]{hyperref}
\hypersetup{colorlinks=true,citecolor=blue,linkcolor=red,%
	urlcolor=blue,pdfstartview=FitH,bookmarksopen=true}
\makeatother

\newcommand{\Rmnum}[1]{\expandafter\@slowromancap\romannumeral #1@}
\newcommand{\ketbra}[2]{\vert #1 \rangle \langle #2 \vert}
\newcommand{\ket}[1]{\vert #1 \rangle}
\newcommand{\bra}[1]{\langle #1 \vert}
\newcommand{\tr}{\mathrm{Tr}}
\newcommand{\I}{\mathrm{i}}

\usepackage{xcolor}

\begin{document}
	\title{Reliable experimental certification of one-way Einstein-Podolsky-Rosen steering}
	
	\author{Qiang Zeng}
	\affiliation{Key Laboratory of Advanced Optoelectronic Quantum Architecture and Measurement of
		Ministry of Education, School of Physics, Beijing Institute of Technology, Beijing 100081, China}
	\affiliation{Centre for Quantum Computation and Communication Technology (Australian Research Council),
		Centre for Quantum Dynamics, Griffith University, Brisbane, Queensland 4111, Australia}
	
	\author{Jiangwei Shang}
	\email{jiangwei.shang@bit.edu.cn}
	\affiliation{Key Laboratory of Advanced Optoelectronic Quantum Architecture and Measurement of
		Ministry of Education, School of Physics, Beijing Institute of Technology, Beijing 100081, China}
	
	\author{H. Chau Nguyen}
	\email{chau.nguyen@uni-siegen.de}
	\affiliation{Naturwissenschaftlich-Technische Fakult\"at, Universit\"at Siegen, 57068 Siegen, Germany}
	
	\author{Xiangdong Zhang}
	\email{zhangxd@bit.edu.cn}
	\affiliation{Key Laboratory of Advanced Optoelectronic Quantum Architecture and Measurement of
		Ministry of Education, School of Physics, Beijing Institute of Technology, Beijing 100081, China}

	\date{\today}

\begin{abstract}
	Quantum steering is a recently-defined form of quantum correlation which lies at the heart of quantum mechanics.
	In difference from other types of quantum correlations, quantum steering is	inherently asymmetric, which implies that it could manifest in one direction but not in the opposite direction. 
	This rather peculiar phenomenon, known as one-way steering, have been demonstrated in several experiments, but subtlety remains. 
	In fact all experiments were shown to be ambiguous until a very recent conclusive one, which however made a crucial use of a high dimensional embedding to get around assumptions. 
	This leaves the question open whether the one-way steering phenomenon can be reliably demonstrated in the genuine two-qubit system.
	Here, we report such an experimental demonstration of one-way steering for a family of two-qubit states.
	Our experimental setup and results thus resolve the subtlety caused by fidelity assumption in previous experiments without the need of higher dimensional embedding. 
	Moreover, our work provides a universal method to characterize one-way steering phenomenon for generic two-qubit states.
\end{abstract}
	
	\maketitle
	
\section{Introduction}%
	Back in 1935, Einstein, Podolsky, and Rosen (EPR) \cite{epr} proposed a gedanken experiment to demonstrate
	that quantum mechanics allows for the so-called “spooky action at a distance.”
	To capture the idea of EPR’s work, Schrödinger later coined the word {\it steering}.
	Naively, steering indicates that the choice of measurements on one side of a bipartite system can specify the possible conditional states of the distant other.
	It remained as an informal intuition until Wiseman {\it et~al.} proposed the rigorous and operational definition for steering from the perspective of quantum information theory~\cite{wisemanSteering2007,jones2007,saunders2010}.
		
	As of nowadays, quantum steering~\cite{Uola2020}, 
	revealed after the other two---entanglement~\cite{horodecki2009} and  
	Bell nonlocality~\cite{brunner2014}---is the last member in the family of quantum nonlocal correlations.   
	Among these three nonlocalities, steering is rather special as its definition possesses inherent asymmetry. 
	Namely, in an entangled system, it is possible that one subsystem is able to steer the other yet the reverse is impossible.
	In the language of theoretical information tasks, the steerability of one party to the other is defined as the ability to remotely generate ensembles
	that could not be explained by means of post-processing on local hidden states (LHS) at the other party.
	While one-way steerability means that the role of the two parties cannot be exchanged.
	
	This unidirectional nature, indicating different level of trust between parties in communication, can be utilized as a unique advantage in one-sided device independent scenarios~\cite{branciard2012,skrzypczyk2018,passaro2015Optimal}, subchannel discrimination~\cite{piani2015}, as well as secure quantum teleportation~\cite{reid2013,he2015Secure}.
	
	Given the important applications of one-way steering, considerable attentions have been devoted to it in recent years~\cite{cavalcantiE2013,chenAll2013,evans2014,cavalcantiD2015,baker2018}. However, as compared to two-way steering, certification of this curious asymmetry of steerability is in fact highly challenging, even purely in theory~\cite{bowlesOneway2014,bowlesSufficient2016,zhu2016,skr2014,quintino2015,nguyengeometry2019,bakerN2020}.
	To prove the steering in one direction, it is natural to use one of various steering inequalities as the criterion, which are essentially focusing on finding at least one specific choice of measurements that excludes any LHS model.
	However, it is tricky to reversely demonstrate the existence of LHS model for all possible choices of measurements given the same state, thus assumptions are often committed.  
	
	First in the Gaussian regime~\cite{reid1989,midgley2010}, asymmetry of steering was observed~\cite{wagner2008} and finally concluded~\cite{handchen2012}.
	However, Gaussian measurements, as a restricted class of measurements, cannot fully characterize one-way steering.
	Indeed, there are explicit examples showing that one-way steerability is destroyed under certain well-chosen non-Gaussian measurements~\cite{chen2002,Wollmann2016}.
	Then some following experiments have assumed two-setting~\cite{sun2016} and multisetting~\cite{xiao2017} projective measurements.
	However, finite-setting measurements still cannot completely reveal LHS models and certify unsteerability.
	
	Overcoming the restrictions on measurement settings, one-way steering have been demonstrated by implementing arbitrary measurements, but still left the fidelity assumption to particular quantum state~\cite{Wollmann2016}. 
	Under this assumption, it is assumed that the targeted states are indeed achieved in the experiment based on the high fidelity obtained.
	But in reality, the actual states in the experiment can easily deviate significantly from the targeted state in the high dimensional space, which leads to the experimental states becoming actually two-way steerable or two-way unsteerable~\cite{Tischler2018a}; see also cases in Appendix~\ref{app:A}.
		
	In the latest one-way steering demonstration~\cite{Tischler2018a}, the authors managed to eliminate both the restrictions on measurement settings and the fidelity assumption, thus for the first time conclusively demonstrated one-way steering. 
	It is however essential to notice that the experimental states, like the one in Ref.~\cite{Wollmann2016}, is realized using a lossy channel.
	Effectively, this corresponds to embedding a symmetric two-qubit Werner state into a qubit-qutrit space in an asymmetric manner. 
	Even though the extra dimension is employed in a trivial way, i.e., as the vacuum-subspace rather than photon-subspace, the need of the third dimension is crucial. 
	This raises the question: can one-way steerability  be realized intrinsically in a genuine two-qubit system? 
	 
	Here we address this question.
	We formulate a family of states in qubit-qubit form, and characterize its one-way steering area accordingly. 
	We experimentally construct the formulated states and rigorously verify the steerability or unsteerability in both directions.
	Our certification results thus resolve the ambiguities caused by fidelity assumption in previous demonstrations without invoking additional dimension to the system.
	It is also worth noting that both our experimental setup and theoretical analysis are adaptive to generic two-qubit states, therefore the method we are presenting in this work indeed provides a universal way of characterizing one-way steerability for two-qubit systems.

\section{The targeted states}%
	Consider a bipartite state $\rho$ shared by two parties $A$ and $B$.
	To introduce the asymmetry, in this work we consider a family of states which mix the singlet state with a designed product state, i.e.,
	\begin{equation}\label{eq:axialstate}
	\rho=p\ket{\Psi^{+}}\bra{\Psi^{+}}+(1-p)\rho_{r}\otimes\frac{\mathbb{I}_B}{2}\,,
	\end{equation}
	where $\ket{\Psi^{+}}=\left(\ket{{0,1}}+\ket{{1,0}}\right)/{\sqrt{2}}$, and $\rho_{r}=\frac{1}{2}\left(\mathbb{I}_A+r\sigma_{z}\right)$ with $\sigma_z$ being the Pauli-Z operator.
	The free parameters $p$ and $r$ can take values from 0 to 1.
	
	To quantify the steerability from A to B of the state, one defines
	\begin{equation}
	\mathcal{R}_{AB}(\rho) = \max_x \{x \ge 0:  \mbox{$\rho^{(x)}$ is unsteerable} \},
	\end{equation}
	where $\rho^{(x)}= x \rho + (1-x) (\mathbb{I}_A \otimes \rho_B)/2$ and $\rho_B=\tr_A(\rho)$ is the reduced state for party B.
	One defines $\mathcal{R}_{BA}(\rho)$ in the similar way.
	It is clear that $\mathcal{R}_{AB}$ and $\mathcal{R}_{BA}$ fully characterize the steerability of $\rho$: $\rho$ is steerable from A to B (or from B to A) if and only if $\mathcal{R}_{AB} (\rho) < 1$ ($\mathcal{R}_{BA} (\rho) < 1$, respectively.)
	It is not obvious that $\mathcal{R}_{AB}(\rho)$ can be computed from its definition.
	However, in Ref.~\cite{nguyengeometry2019}, the quantity $\mathcal{R}_{AB}(\rho)$ is given a geometrical meaning, implied in its name as critical radius of steering, and an algorithm to compute its value for an arbitrary state is also given.
		
\begin{figure}[t]
	\includegraphics[width=.65\columnwidth]{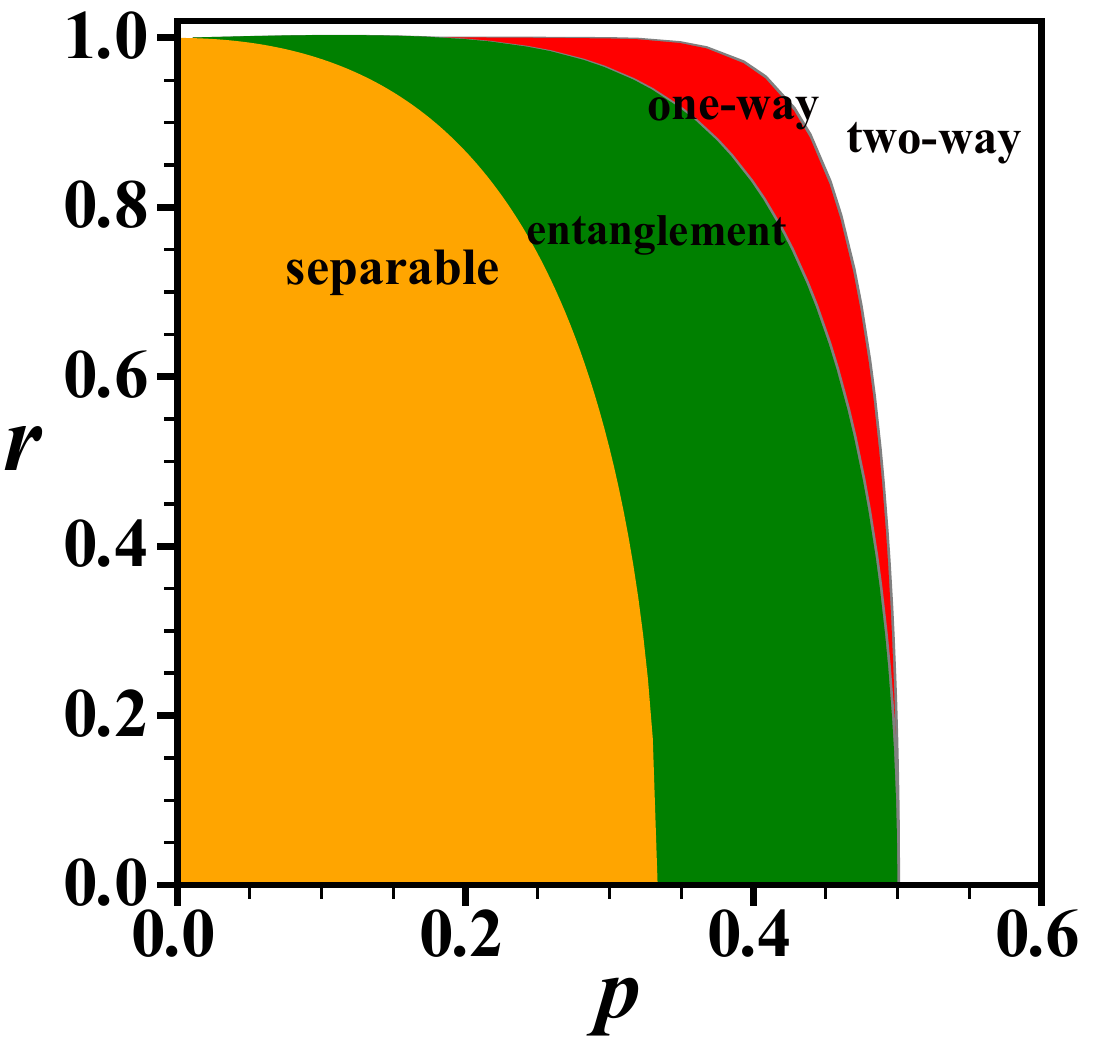}
	\caption{\label{fig:steerArea} 
		Parametric areas of steering hierarchies for the two-parameter targeted states as defined in Eq.~\eqref{eq:axialstate}.
		The orange area indicates the corresponding states are separable, while the two-way unsteerable states (also known as entanglement) lie in the green area. The pure white area includes all two-way steerable states.
		The one-way steering area is colored with red, which is clearly smaller as compared to other areas. 
	}
\end{figure}

	In fact, states in the form of Eq.~\eqref{eq:axialstate} carry an axial symmetry~\cite{nguyengeometry2019}, which allows for an easy characterization of the one-way steerable areas as illustrated in Fig.~\ref{fig:steerArea}.
	One-way steering is in fact very delicate. 
	It only manifests in a small parametric area comparing to other hierarchies. 
	In Fig.~\ref{fig:steerArea}, the narrowness of one-way steerable area is well exhibited as the red area is much smaller than other areas. 
	Because of this narrowness, we note that precise parameter manipulation and effective noise control are highly demanding in the experimental demonstration.

\begin{figure*}[htbp]%
	\includegraphics[width=1.45\columnwidth]{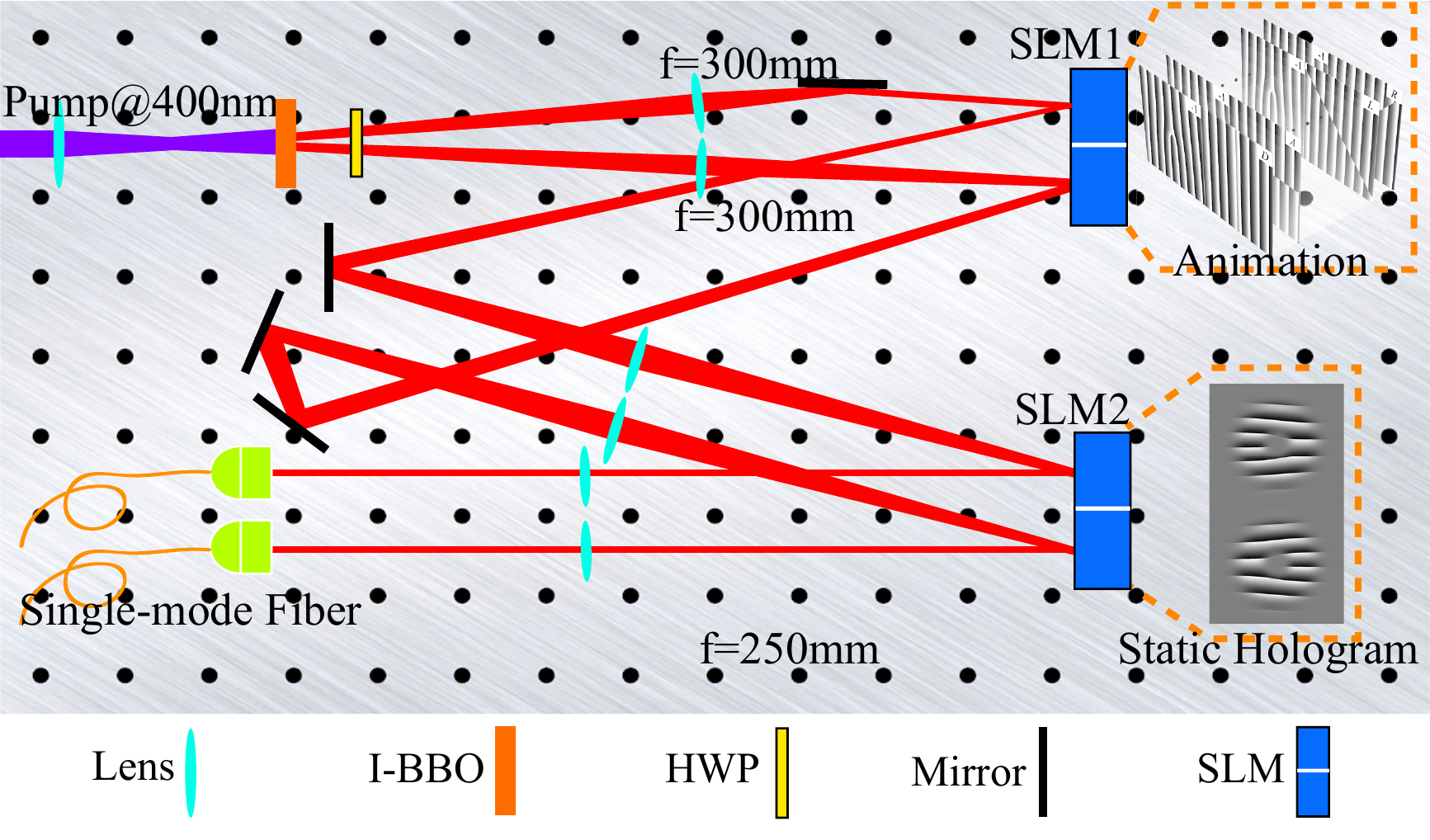}
	\caption{\label{fig:1ws_sketch} 
		Sketch of the experimental platform.
		Pumper: Coherent Chameleon (centered at 800~nm).
		HWP: half-wave plate.
		SLM: spatial light modulator.
		Collector: photon to fiber collector embedded with narrow bandwidth filter (800~nm~$\pm$ 5~nm).
		SMF: single-mode fiber with diameter of 5~$\mu$m.
		A standard type-I spontaneous parametric down-conversion (SPDC) with the efficiency being around 15\% is employed to generate the OAM entanglement.
		In order to obtain as many as entangled photon pairs without losing high-order modes, a non-linear type-I BBO crystal with medium thickness (3~mm) is adopted.
		As for the animation configuration, in each path, the size of the hologram is $960\times1080$ (pixel) with the exposure time being 0.02~s.
		We set the number of sampling time as $N=800$ to guarantee that every hologram is effectively estimated even when the $p$ and $r$ are relatively small.
		The coincidence accumulation time is set to 20~s to ensure that the animation is fully applied.
		Each of the two SLMs has a loss of photons due to the optical grating, whose efficiency is 49.6\%. The efficiency of the collection system is 70.1\%. Thus the overall efficiency of our experiment setup is about 17.2\%.
	}
\end{figure*}
%

\section{Animation preparation}%
	To construct the above targeted states, we need to obtain the two involved components and mix them in a natural way. 
	It is proposed recently that employing spatial light modulator (SLM) loading animation can introduce tunable isotropic noise into the maximally entangled states, and construct the isotropic states~\cite{zeng2018}. 
	This setup can be utilized to characterize high dimensional steering, or even the underlying noise suppression phenomenon, for two-setting mutually unbiased bases (MUBs)~\cite{zeng2018} and further $N$-setting MUBs~\cite{qu2021}. 
		
	The basic idea of this setup is to decompose the noise state into the representation of MUBs which are to be measured, and generate the corresponding elements in this decomposition stochastically, thus mimic the noise state in a statistical way. 
	For example, if we need a white noise $\mathbb{I}/2$ on one party given the measurement bases are $\ket{0}$, $\ket{1}$, $\ket{+}$ and $\ket{-}$, where $\ket{\pm}=\ket{0}\pm\ket{1}$, we can simply decompose 
	$$\frac{\mathbb{I}}{2}=\frac{1}{4}\left( \ketbra{0}{0}+\ketbra{1}{1}+\ketbra{+}{+}+\ketbra{-}{-} \right).$$
	Then we generate the computer-generated holograms (CGH) which correspond to these decomposed elements, to form a random pool and sample from this pool with designated probabilities to construct the animation. 
	Note that albeit this setup is encoded with orbital angular momentum (OAM) modes~\cite{zeng2018,qu2021}, which is widely applicable in quantum information field~\cite{karimiEfficient2009,*karimiSpin2010,*slussarenko2011,*zeng2016}, 
	it is also applicable for encoding with general spatial modes of photons~\cite{pixel2020} or even electrons~\cite{grillo2017Observation}.
	
	In our experiment, the qubit is encoded with the
	OAM modes ${l=\pm 2}$, namely, $\ket{0}\equiv\ket{{l=-2}}$ and $\ket{1}\equiv\ket{{l=2}}$. 
	Then, an over-complete tomography scheme \cite{PhysRevA.84.062101} with six
	measurement outcomes on each side is
	implemented to characterize the targeted states.
	Specifically, the six measurement outcomes
	are written as $\ket{X_{m}}=\ket{{0}}+({-1})^{m}\ket{1}$, $\ket{Y_{n}}=\ket{{0}}+\I({-1})^{n}\ket{1}$, $\ket{Z_{0}}=\ket{{0}}$, and $\ket{Z_{1}}=\ket{1}$,
	with $m,n\in\{0,1\}$.
	Based on these expressions, the maximally mixed state on one party can be decomposed as \cite{zeng2018}
	\begin{equation}\label{eq:I/2}
	\frac{\mathbb{I}}{2}=\frac{1}{6}\left(\sum^{1}_{m=0}\ket{X_{m}}\bra{X_{m}}+\sum^{1}_{n=0}\ket{Y_{n}}\bra{Y_{n}}+\sum^{1}_{l=0}\ket{Z_{l}}\bra{Z_{l}}
	\right).
	\end{equation}
	In addition, the designed product component in the targeted state can be expressed as
	\begin{align}
	\rho_{r}\otimes\frac{\mathbb{I}}{2}&=\left[r\ketbra{0}{0}+(1-r)\frac{\mathbb{I}}{2}\right]\otimes\frac{\mathbb{I}}{2}	\nonumber	\\ 
	&=r\ketbra{0}{0}\otimes\frac{\mathbb{I}}{2}+(1-r)\frac{\mathbb{I}}{2}\otimes\frac{\mathbb{I}}{2}\label{eq:rhor},
	\end{align}
	from which we can easily see that the second term in Eq.~\eqref{eq:rhor} correspond to 36 different two-qubit product states according to Eq.~\eqref{eq:I/2}, which requires 36 corresponding holograms being in the random pool.
	It is also obvious that the second term essentially represents an isotropic noise. 
	Similarly, the first term in Eq.~\eqref{eq:rhor} requires 6 different holograms being in the random pool, which represents the product state of a pure state $\ket{0}$ on Alice's side and a maximally mixed state on Bob's side.
		
	Based on the above decomposition, we generate the CGHs and form the random pool accordingly. 
	Then we sample from the random pool using random number generator to obtain the targeted animations~\cite{zeng2018}.
	The sampling rule, which specifies the exact probability for each hologram given particular $p$ and $r$, is provided in Appendix~\ref{app:B}.
	A sketch of the experiment platform is shown in Fig.~\ref{fig:1ws_sketch}, and the detailed description on the implementation as well as the real scale version of our setup is provided in Appendix~\ref{app:C}.
	
	As we have obtained the required animations for any given parameter settings $p$ and $r$, what we need to do next is to check whether the final experimental states generated by the animations are indeed one-way steerable.
	Moreover, we have to characterize the correspondence between the theoretical predictions and the final experimental results.

\begin{figure}[t]
	\includegraphics[width=\columnwidth]{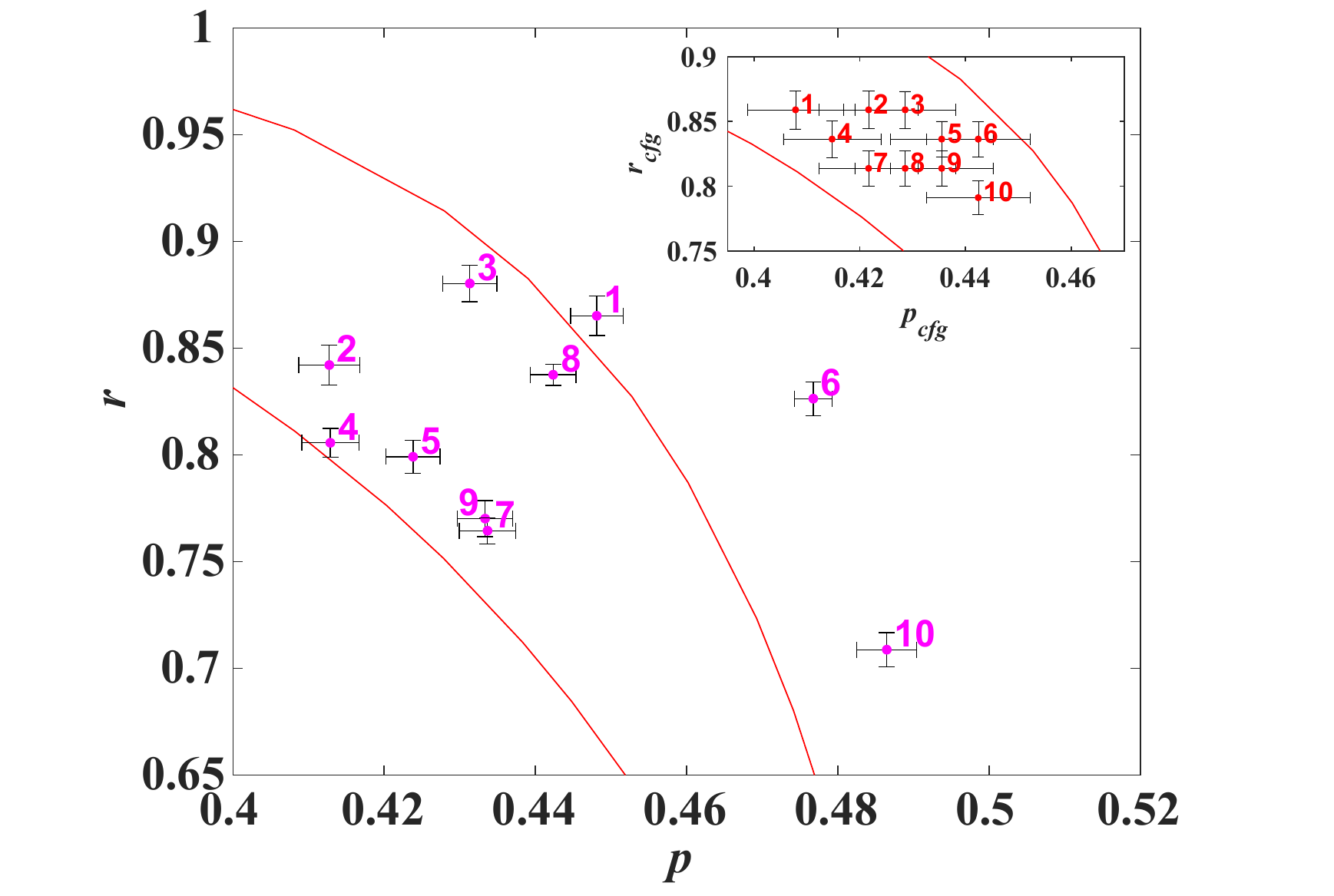}
	\caption{\label{fig:config}
		Distribution of experimentally reconstructed states. 
		Parameters $p$ and $r$ are retrieved from the reconstructed density matrices, which are denoted by the magenta dots. 
		For comparison, the inset presents the parameter configuration $p_{\text{cfg}}$ and $r_{\text{cfg}}$ of the ten targeted states, which are denoted by the red dots. 
		The uncertainties of $p$ and $r$ are obtained from the bootstrapping simulation with 20 variations on each of the experimental over-complete tomography results. 
		The uncertainties of $p_{\text{cfg}}$ and $r_{\text{cfg}}$ are induced by the inherent nonequilibrium of photon numbers between the maximally entangled state and the product state; see Appendix~\ref{app:devi} for details.
	}
\end{figure}
%

\section{Results}%
	With the above experimental platform, we select ten targeted states with configured parameters $p_{\text{cfg}}$ and $r_{\text{cfg}}$.
	The distribution of these ten states are exhibited in the embedded inset in Fig.~\ref{fig:config}.
	Note that the configured parameters $p_{\text{cfg}}$ and $r_{\text{cfg}}$ which regularly distribute in the predicted one-way steerable area are merely the target for the task. 
	We make no assumption on the final experimental states; they are not assumed to be even of the form of Eq.~\eqref{eq:axialstate}.
	The computation of the critical radius consider the input as an arbitrary state.
	
	With ten targeted animations loading on SLM 1 respectively, we implement over-complete quantum state tomography on SLM 2 
	(see Appendix~\ref{app:tomo} for detailed results)
	to reconstruct the experimental states, based on whose density matrices, we then compute the corresponding critical radii $\mathcal{R}_{AB}$ and $\mathcal{R}_{BA}$.	
	
\begin{figure}[t]
	\includegraphics[width=\columnwidth]{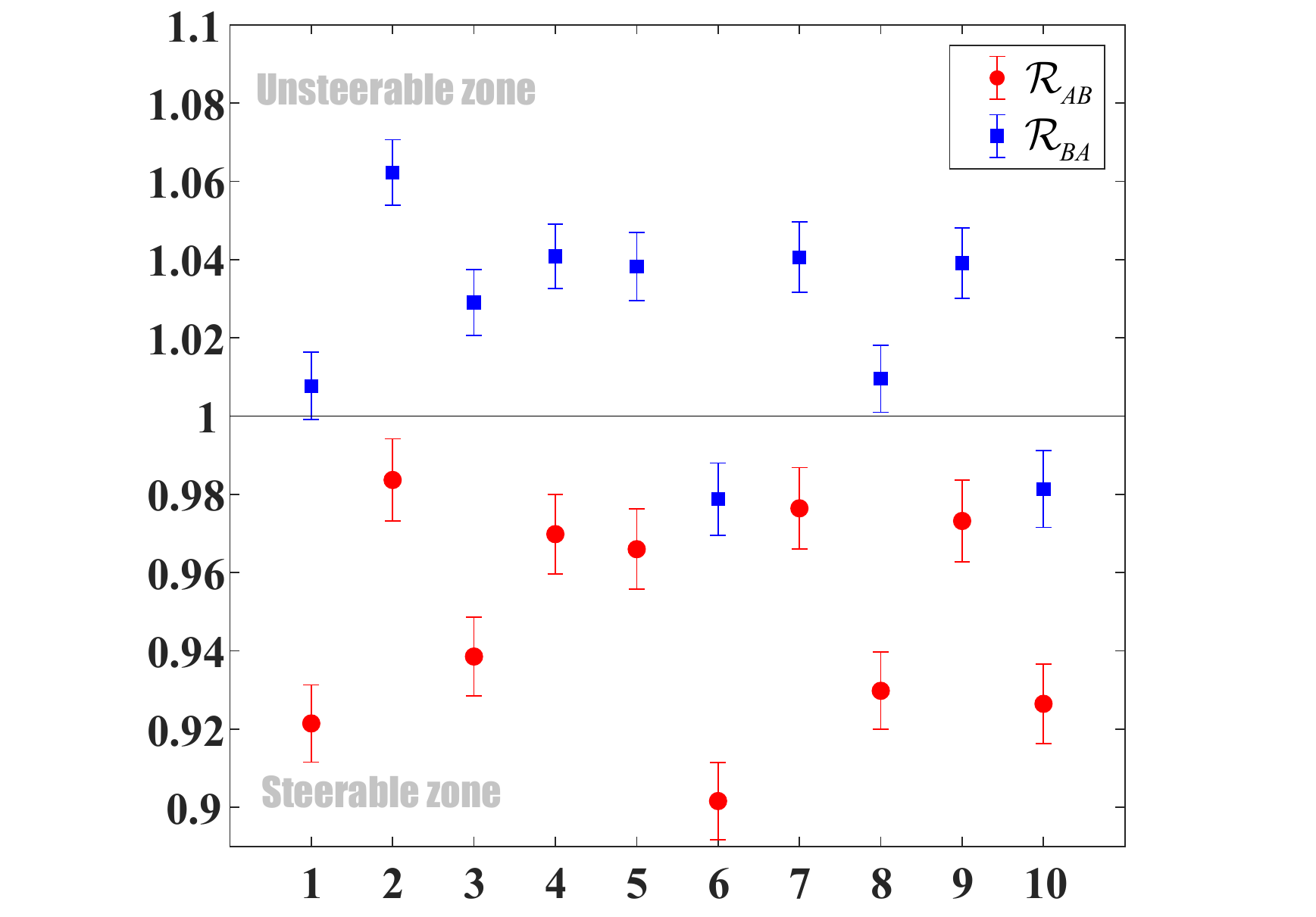}
	\caption{\label{fig:results} Experimental results of one-way EPR steering certification. Numbers 1 to 10 denote the ten obtained states. Red dots refer to the steering measure of Alice steering Bob, and blue squares refer to the measure in the opposite direction.}
\end{figure}

	In Fig.~\ref{fig:results}, we present our one-way steering certification results, where all the critical radii of the selected ten states in both directions are exhibited.
	The red dots and blue squares refer to the steering direction from $A$ to $B$, and from $B$ to $A$, respectively.
	As we mentioned before, one can conclude steerability in certain direction if and only if the corresponding critical radius is less than 1.
	Clearly, for most states except 1, 6 and 10, their critical radii of two opposite directions distinctly lie in the separate zones, which clearly certifies the one-way steerability.
	For state 1, its critical radius from Bob to Alice $\mathcal{R}_{BA}$ almost certify the unsteerability but the error bar crosses the boundary.
	For states 6 and 10, both $\mathcal{R}_{AB}$ and $\mathcal{R}_{BA}$ are less than 1, which indicates two-way steerable.

	To find out why we fail in demonstrating one-way steering with states 1, 6 and 10, we retrieve the closest $p$ and $r$ to high accuracy from their reconstructed density matrices. 
	We present the distribution of retrieved $p$ and $r$ in Fig.~\ref{fig:config}, from which 
	we see a significant deviation of the states 6 and 10 from the configured parameters, also the predicted one-way steering area, while state 1 lies outside but close to the one-way steerable area, far from the targets.   
	On the contrary, all other states which are rigorously certified to be one-way steerable lie in the one-way steering area as expected. 
	Combining Figs.~\ref{fig:config} and \ref{fig:results}, we actually demonstrate the correspondence between the theoretical predictions and the final experimental results, that is, all certified states indeed lie in one-way area, while false one-way steering states are unambiguously excluded, despite the fact those states are with high fidelity (the detailed values of fidelity are provided in 
	Appendix~\ref{app:devi}).

	The deviation of states 1, 6 and 10 in fact shows that random errors can be significantly enhanced in one-way steering demonstration.
	We note that the considered parameters of the states are only two out of fifteen independent parameters of a two-qubit states, while
	experimental deviations present not only in the two but
	all fifteen parameters. 
	If one only concerns 2 specific parameters, the deviations in other parameters are very likely
	to cause the states shifting to two-way steerable or two-way unsteerable, despite the fact that the overall fidelity is high.
	
	The random errors are known coming from the inherent fluctuation of OAM modes generated in the SPDC process by the BBO crystal, and from the inevitable shot noise.
	We have quantified the possible fluctuation induced by the BBO crystal, which is exhibited with the error bar of the configured parameter in the inset of Fig.~\ref{fig:config} (see also the value of error bars in 
	Appendix~\ref{app:devi}).
	Nevertheless, for the extraordinary deviations such as states 6 and 10, we also attribute these cases partially to accidental disturbance such as wavelength shift of the pumper and note that increasing the trials of experiment states can effectively eliminate the impact of accidental disturbance.
	We note that using specific parameters to select targeted states is helpful, but certification of one-way steerability must not make any assumption on the form of experimental state,
	thus our results support that drawing conclusion based on high fidelity is indeed problematic~\cite{Tischler2018a}. 
	
\section{Conclusion}%
	We have reported an experimental certification of one-way EPR steering for a family of natural two-qubit states, in which all common assumptions on state fidelity or measurement settings have been eliminated. 
	Our experimental setup possesses high level of precision and manipulation of parameters. With the fidelity assumption eliminated, we have conclusively demonstrated one-way steering, and unambiguously excluded false one-way steering cases caused by random errors, thus presented reliable results without invoking the dimensional asymmetry.
	The involved qubits are encoded using orbital angular momentum of photons, and the experimental states are achieved by employing spatial light modulator by loading the designed animations. 
				
	Several direct extensions of this work is promising. 
	One is to characterize one-way steering effect in higher dimensional systems, which is technically achievable since the orbital angular momentum degree of freedom is naturally with high dimensionality. 
	Other extensions could be finding and verifying optimal one-way steerable states that are with largest possible one-way steering area, since the theory as well as the experimental platform in this work are adaptive to generic two-qubit states.

\acknowledgments
	We are grateful to Howard Wiseman, Chuan-Feng Li, and Ya Xiao for helpful discussions.
	We especially thank the authors of Ref.~\cite{xiao2017} for providing us with their experiment data and useful discussions.
	This work was supported by the National Key R\&D Program of China
	under Grant No.~2017YFA0303800 and the National Natural Science Foundation of
	China through Grant Nos.~11574031, 61421001, and 11805010.
	Q.Z. acknowledges support by the Office of China Postdoctoral Council The International Postdoctoral Exchange Fellowship Program (Grant No. 20190096).
	J.S. also acknowledges support by the Beijing Institute of Technology Research Fund Program for Young Scholars.
	H.C.N. is supported by the Deutsche Forschungsgemeinschaft (DFG, German Research Foundation - 447948357) and the ERC (Consolidator Grant 683107/TempoQ).


\appendix

\section{Ambiguity in concluding one-way steering from previous two-qubit demonstration}\label{app:A}
	As we mentioned in the main text, except for the recent experiment reported in Ref.~\cite{Tischler2018a}, which had fully eliminated the assumptions on fidelity and measurement settings, other existing one-way steering experiments are proved to be inconclusive. This issue in fact has been discussed in Ref.~\cite{Tischler2018a}. However the implementation in Ref.~\cite{Tischler2018a} involves a lossy channel, which raises subtlety in the asymmetric dimensionality. Thus within natural qubit-qubit systems, the one-way steering experiments still remain inconclusive. 

	To illustrate this issue, we take the experiment reported in Ref.~\cite{xiao2017} as an example.
	The state under consideration is
\begin{equation*}
	\rho = p \ketbra{\theta}{\theta} + (1-p) \mathbb{I}_A \otimes \rho_B,
\end{equation*}
	where $\ket{\theta} = \cos \theta \ket{00} + \sin \theta \ket{11}$ and $\rho_B = \mathrm{Tr}_A [\ketbra{\theta}{\theta}]$. 
	It has been shown in Ref.~\cite{bowlesSufficient2016} that for this set of states, $p>1/2$ and $\cos^2 (2 \theta) \ge (2 p -1)/[(2 -p)p^3]$ lead to the case where the states are steerable from $A$ to $B$, yet unsteerable from $B$ to $A$. 

\begin{figure}[ht]
	\centering
	\includegraphics[width=1.1\columnwidth]{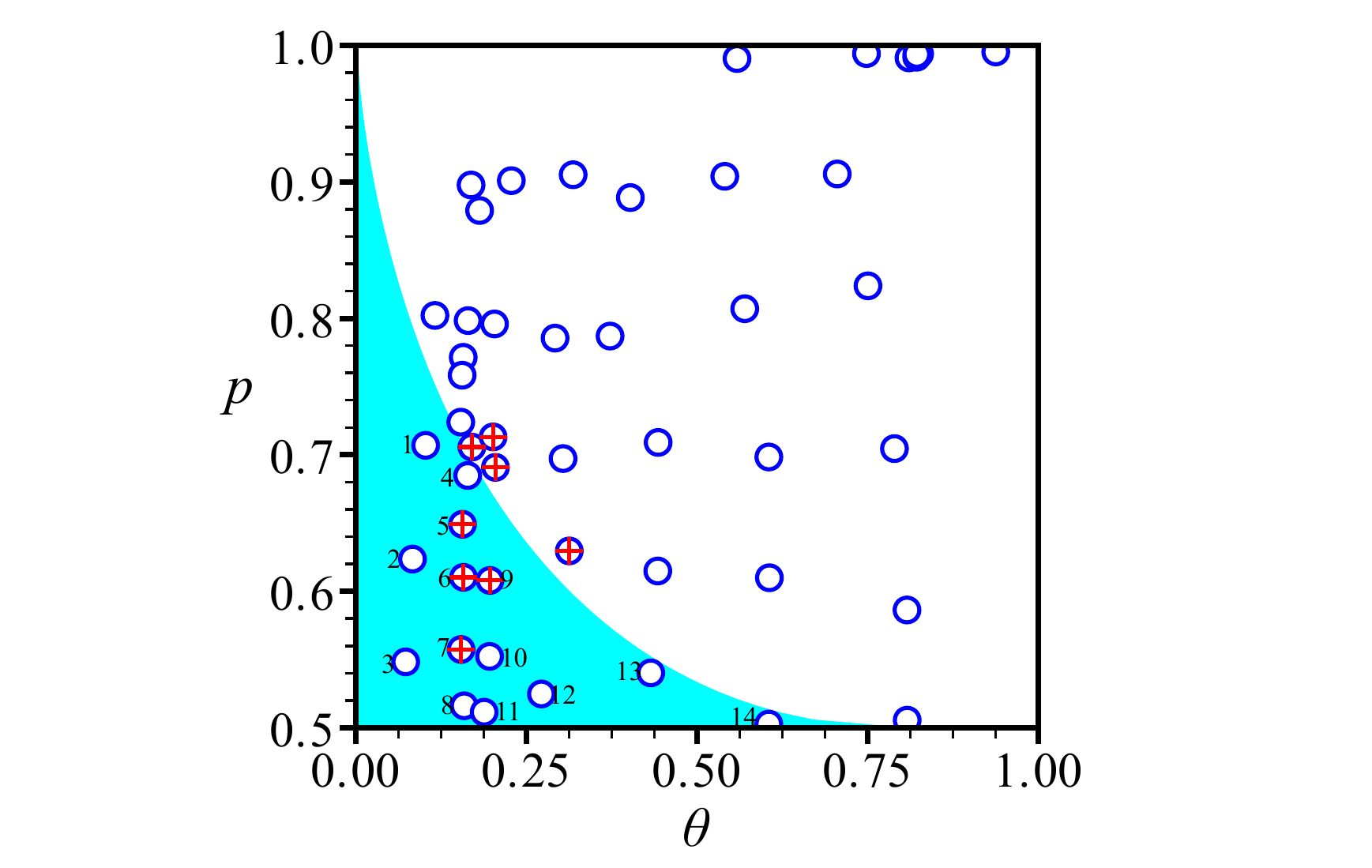}
	\caption{\label{fig:xiao} Reconstructed states and the distribution of retrieved parameters according to the raw data from Ref.~\cite{xiao2017}. Out of $14$ states (numbered beside the circle) which have the retrieved parameters in the predicted one-way steering area (cyan), only $4$ of them are proved to be unsteerable (denoted with red cross). Note that the indicated one-way steering area is only an inner bound of the true one-way steering area, thus there are also $4$ states outside this area which turn out to be one-way steerable.}
\end{figure}
\begin{figure*}[t]%
	\includegraphics[width=0.7\textwidth]{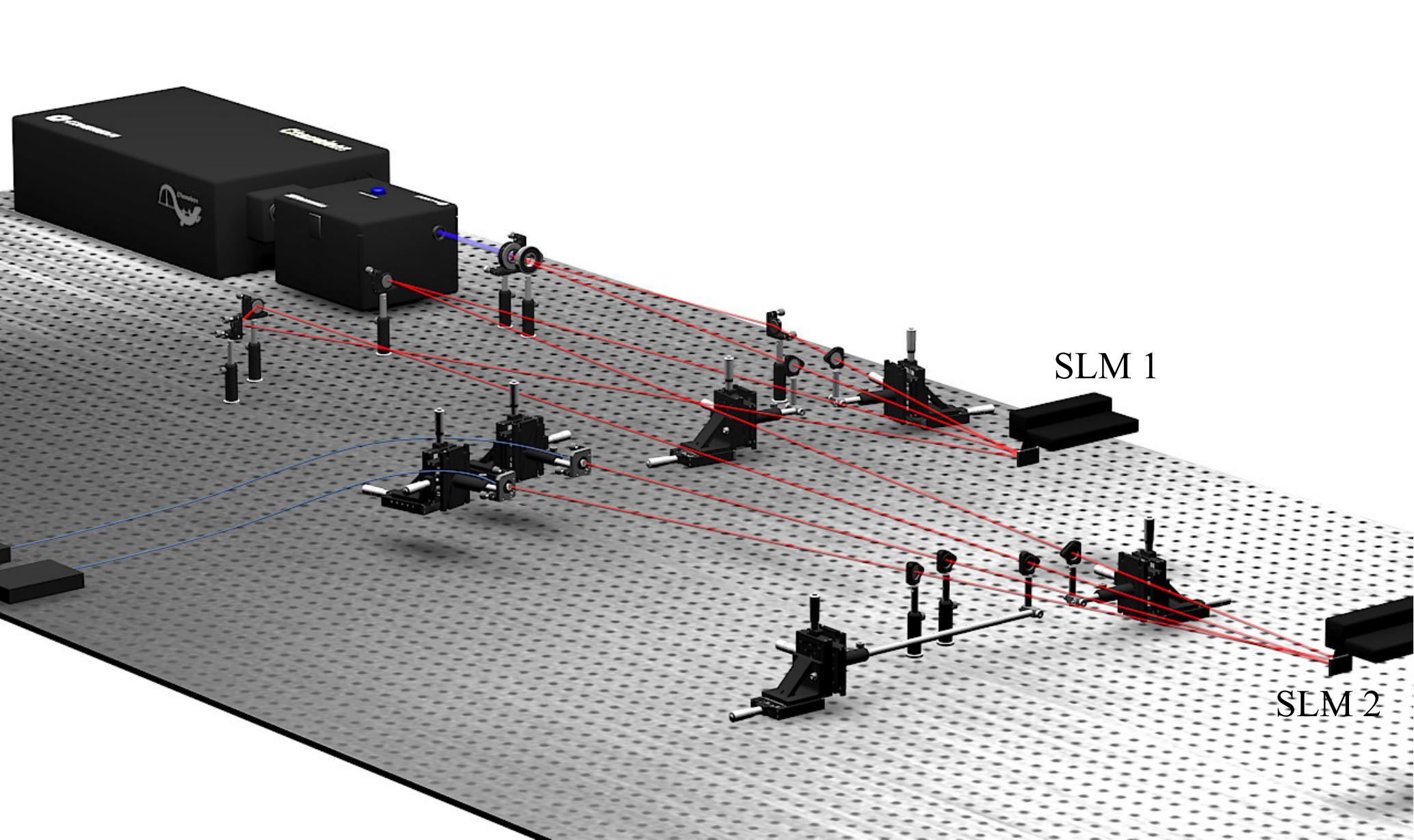}
	\caption{\label{fig:sp} Real scale experimental platform. The supports of spatial light modulators are omitted for better viewing.}
\end{figure*}

	Although the criterion for one-way steering adopted in this experiment is limited with finite number of measurement settings,
	their data however contain the part which is available for the quantum state tomography.
	With the experimental data obtained in Ref.~\cite{xiao2017}, we then reconstruct the $50$ experimental states. 
	The corresponding parameters $\theta$ and $p$ are retrieved to a high accuracy. 
	Here we note that in their practical data processing, the parameters $\theta$ and $p$ are directly obtained from the readout of the experimental devices, where the high fidelity assumption is committed. 
	In the following we can see this assumption is not as reliable as expected.
	
	In Fig~\ref{fig:xiao}, we present the reconstruction and certification results. All those $50$ states are exhibited denoting with blue circles. 
	The predicted one-way steering parametric area is painted with cyan color. Moreover, those states that are indeed one-way steerable certified by the critical radius theory are highlighted with red cross.
	From Fig~\ref{fig:xiao}, we can see that out of all $50$ states, $14$ of them have the retrieved parameters falling into the one-way steerable area. 
	However, the claim that all these states are one-way steerable is in fact incorrect, for there are only $4$ of them proved to be unsteerable in the designated direction.
	
	
%
\begin{figure*}[t]%
	\includegraphics[width=.9\textwidth]{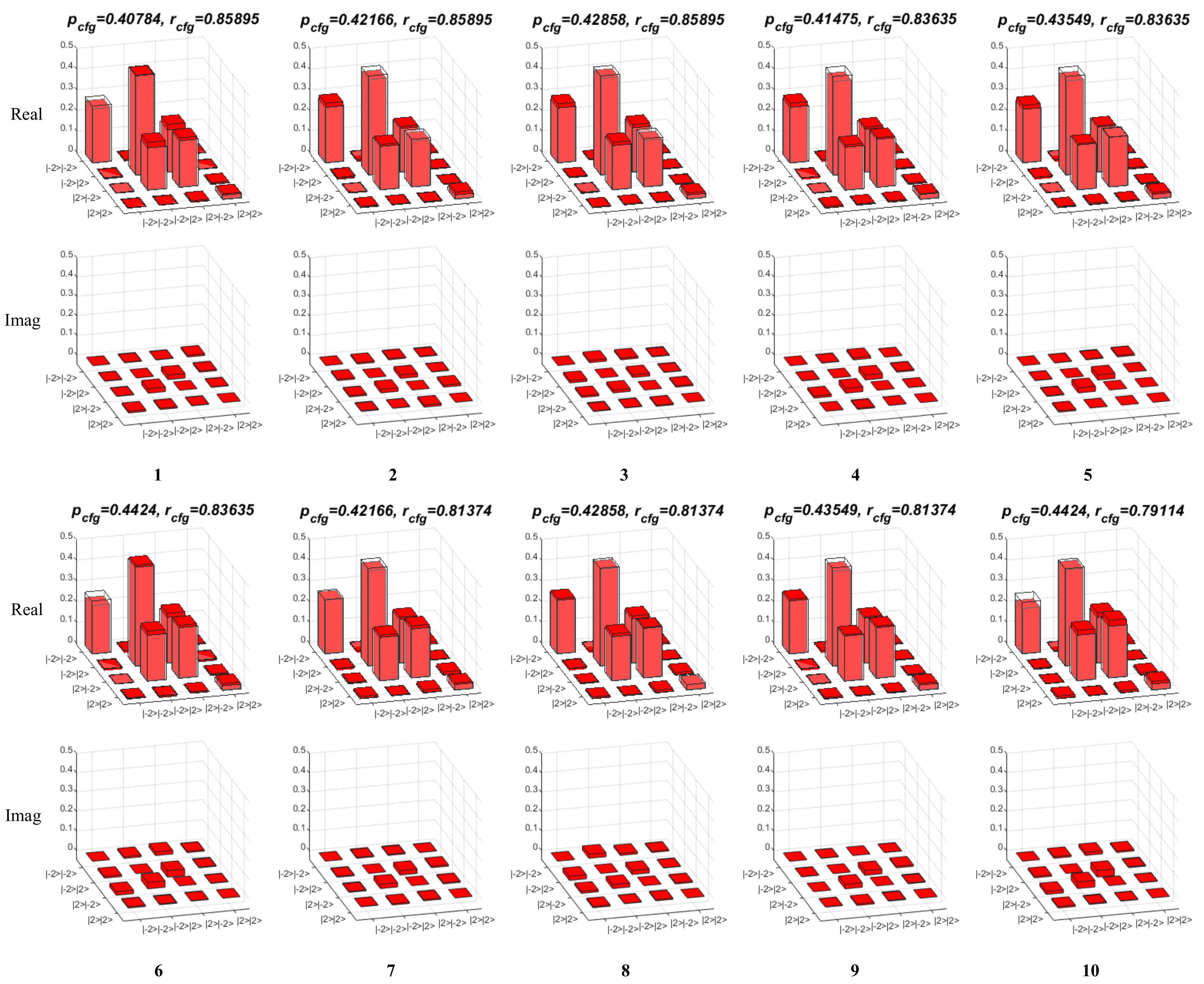}
	\caption{\label{fig:tomo} 
		Reconstructed density matrices of the ten targeted states. 
		Because of the narrowness of the one-way steering area, the parameters for the selected one-way steering states are very similar to each other. 		
		To highlight the contrast, we use solid red bars to refer to the density matrices of the experimentally reconstructed states, and transparent wire-frame bars to refer to the density matrices of the configured target states.
		In this setting, we can also easily distinguish the deviated states 1, 6 and 10 from other states as their solid bars have relatively larger gap as compared to the transparent bars. 
		It is worth noting that these gaps are barely visible purely from the entry values of the density matrices since all the reconstructed states have very high fidelity.
	}
\end{figure*}
%

\section{Sampling rules for constructing animation}\label{app:B}	
	We denote the sample probability of the isotropic part as $\{p_i\}_{36}$.
	Besides, we denote the emerging probability of the six holograms which correspond to the asymmetric product state as $\{p_s\}_6$.
	Also, an extra hologram with no mode modulation is added to represent the maximally entangled state, whose probability is denoted as $\{p_e\}$.
	
	Since the exposure time $t$ of each hologram is fixed, to calculate the emerging probabilities of different holograms for given $p$ and $r$, we need to consider a certain time interval $T$. 
	During this time interval, the hologram corresponding to the maximally entangled state shall emerge $pT/t$ times, while the product state shall emerge $(1-p)\left[r+2(1-r)\right]T/t$ times, so that the relative probability between the maximally entangled state and the product state is $$\eta_e=\frac{p}{(1-p)\left[r+2(1-r)\right]},$$ which leads to $p_ e=p/\left[(1-p)(2-r)+p\right]$.
	Therefore, the probability of the designed product part which we denote as $p_p$ is given by $p_p=1-p_e$.

	Likewise, within the designed product state, the relative emerging probability $\eta_p$ between the pure state part and the isotropic part is $$\eta_p=\frac{r}{2(1-r)}.$$ Given there are 6 $p_s$ and 36 $p_i$, we finally obtain $p_s=\eta_p{p_p}/6$ and $p_i=(1-\eta_p)p_p/36$.
	
	In practice, the sampling procedure involves four random generators, which we denote as $\lambda_{1,2,3,4}$.
	If $\lambda_1\le{p_e}$, the hologram corresponding to the maximally entangled state will be selected and added as a frame to the animation.  
	Otherwise if $\lambda_1 > {p_e}$, which indicates the designed product state, the second random generator $\lambda_2$ is needed to determine whether the pure state or the isotropic part is selected. 
	In this case, if $\lambda_2\le{\eta}$ and $\lambda_{3}\in\frac1{6}[\alpha-1,\alpha]$ where $\alpha\in\{1,2,...,6\}$, the $\alpha\text{-th}$ hologram of the pure state part will be selected. 
	However, if $\lambda_2 > {\eta}$, and $\lambda_{4}\in\frac1{36}[\beta-1,\beta]$ where $\beta\in\{1,2,...,36\}$, the $\beta\text{-th}$ hologram of the isotropic part will be selected.  
	After several rounds of sampling, we can eventually construct an animation that is corresponding to the specific targeted state. 
	
	It is worth stressing that although each frame in the animation is fixed, the random sampling ensures that it is impossible to predict which of these frames would be applied to the incident photon, thus the overall effect of the animation indeed mimics the stochastic process statistically.

\section{Experimental implementation and real scale experimental platform}\label{app:C}	
	In Fig.~\ref{fig:sp}, we show the real scale schematic of our experimental platform. 
	In order to obtain the maximally entangled OAM state $\ket{\Psi^{+}}$, we utilize a pulsed laser with pulse length of 1~fs to impact a type-I BBO crystal.
	The pulsed laser centres at 400~nm with the power of 180~mw.
	A half-wave plate (HWP) is inserted to align the photons at horizontal polarization for meeting the application requirement of the spatial light modulator (SLM).
	The subsequent lenses with 300~mm focal length are to image the entangled photons on SLM~1, where an animation is loaded.
	A proper modulating process requires the image size of the incident photons matches with the screen scale of the SLM, so we set the distance between the lenses and SLM~1 as 350~mm, by which we render the image of the incident photons on the subsequent SLM~1 with the image being properly shrunk.
	Likewise, the following lenses and SLM~2 keep the same structure.
	However, instead of using animation, we set static holograms loaded on SLM~2, which is to accomplish the state tomography upon associating with the single-mode fibre.
	Besides, in order to achieve a high efficiency of collecting photons, we insert a pair of lenses with 250~mm focal length in front of the two collectors, where two micro lenses with 7.5~mm focal length are also embedded respectively (not shown in the sketch).

\begin{table*}[t]
	\caption{\label{tab:parameter} Parameter settings of the ten targeted states.}
	\begin{ruledtabular}
		\begin{tabular}{ccccccc}
			&1&2&3&4&5\\
			\hline
			$p_\text{ipt}$& 0.36875&	0.38125	&	0.3875&	0.375&	0.39375		\\
			$r_\text{ipt}$& 0.95	&	0.95	&	0.95&	0.925&	0.925 		\\
			$p_\text{cfg}$& $0.4078\pm	0.0091$	&	$0.4217\pm0.0094$	&	$0.4286\pm0.0095$&	$0.4148\pm0.0092$&	$0.4355\pm0.0097$	\\			
			$r_\text{cfg}$& $0.859\pm0.0234$	&	$0.859\pm0.0234$	&	$0.859\pm0.0234$&	$0.8363\pm0.0228$&	$0.8363\pm0.0228$ 	\\
			Fidelity\footnotemark[1]
			&$0.9964\pm0.0008$	&$0.9956\pm0.0006$	&$0.9957\pm0.0007$	&$0.9960\pm0.0006$	&	$0.9963\pm0.0004$ \\
			\hline\hline
			&6&7&8&9&10\\
			\hline
			$p_\text{ipt}$& 0.4&		0.38125&	0.3875&		0.39375&	0.4		 \\
			$r_\text{ipt}$& 0.925&		0.9&		0.9&		0.9&		0.875	 \\
			$p_\text{cfg}$& $0.4424\pm0.0098$&	$0.4217\pm0.0094$&	$0.4286\pm0.0095$&	$0.4355\pm0.0097$&	$0.4424\pm0.0098$	\\
			$r_\text{cfg}$&$0.8363\pm0.0228$&	$0.8137\pm0.0221$&	$0.8137\pm0.0221$&	$0.8137\pm0.0221$&	$0.7911\pm0.0215$	\\
			Fidelity
			&$0.9960\pm0.0005$	&	$0.9966\pm0.0006$	&	$0.9970\pm0.0005$	&	$0.9958\pm0.0013$	&	$0.9951\pm0.0005$	 \\
		\end{tabular}
	\end{ruledtabular}
	\footnotetext[1]{The fidelity is defined as 	$[\text{Tr}(\sqrt{\sqrt{\rho}\rho_{0}\sqrt{\rho}})]^2$.}
\end{table*}
%

\section{Over-complete quantum state tomography}\label{app:tomo}	
	In this section, we show the tomography results of the ten targeted states by using an over-complete quantum state tomography scheme. Figure~\ref{fig:tomo} shows the reconstructed density matrices.

\section{Parameter configuration and analysis on deviation between the reconstructed states and the targeted states}\label{app:devi}
	In our animation construction, we prepare the involved computer-generated holograms (CGH) in advance and form the random pool. 
	We then select one of those CGHs to append as the frame of the animation according to the sampling rules described in the main text and do the sampling for many times to finally construct the animation.
	In the parameter configurations, the number of photons corresponding to the maximally entangled state and the number of photons which are to be transformed into designed product states are supposed to be equal. 
	However, in practice, due to the inherent unbalanced distribution of the OAM modes under the SPDC process and the different collecting efficiency between these two types of photons, these two numbers are slightly different. 
	
	Specifically, the ratio between these two numbers in our experiment is $\alpha=1.106\pm0.0246$. 
	Consequently, we have $p_\text{cfg}=p_\text{ipt}*\alpha$ and $r_\text{cfg}=r_\text{ipt}/\alpha$, where $p_\text{cfg} (r_\text{cfg})$ is the parameter we aim to achieve when we input $p_\text{ipt} (r_\text{ipt})$. 
	In Table~\ref{tab:parameter}, we list all the parameter settings in detail. 
	The significant fluctuation of the ratio is thus the main cause of the deviation between the obtained and the targeted states. 
	Note that the deviation is also affected by the shot noise and slight crosstalk of the OAM measurement bases. 
	
	Observing the fidelity, we see that even states 6 and 10 which have extraordinary deviations, still keep very high fidelity values, from which we note again that the value of fidelity can work as reference yet drawing conclusions based on high fidelity can be seriously problematic~\cite{Tischler2018a}.


\begin{thebibliography}{43}%
\makeatletter
\providecommand \@ifxundefined [1]{%
 \@ifx{#1\undefined}
}%
\providecommand \@ifnum [1]{%
 \ifnum #1\expandafter \@firstoftwo
 \else \expandafter \@secondoftwo
 \fi
}%
\providecommand \@ifx [1]{%
 \ifx #1\expandafter \@firstoftwo
 \else \expandafter \@secondoftwo
 \fi
}%
\providecommand \natexlab [1]{#1}%
\providecommand \enquote  [1]{``#1''}%
\providecommand \bibnamefont  [1]{#1}%
\providecommand \bibfnamefont [1]{#1}%
\providecommand \citenamefont [1]{#1}%
\providecommand \href@noop [0]{\@secondoftwo}%
\providecommand \href [0]{\begingroup \@sanitize@url \@href}%
\providecommand \@href[1]{\@@startlink{#1}\@@href}%
\providecommand \@@href[1]{\endgroup#1\@@endlink}%
\providecommand \@sanitize@url [0]{\catcode `\\12\catcode `\$12\catcode
  `\&12\catcode `\#12\catcode `\^12\catcode `\_12\catcode `\%12\relax}%
\providecommand \@@startlink[1]{}%
\providecommand \@@endlink[0]{}%
\providecommand \url  [0]{\begingroup\@sanitize@url \@url }%
\providecommand \@url [1]{\endgroup\@href {#1}{\urlprefix }}%
\providecommand \urlprefix  [0]{URL }%
\providecommand \Eprint [0]{\href }%
\providecommand \doibase [0]{https://doi.org/}%
\providecommand \selectlanguage [0]{\@gobble}%
\providecommand \bibinfo  [0]{\@secondoftwo}%
\providecommand \bibfield  [0]{\@secondoftwo}%
\providecommand \translation [1]{[#1]}%
\providecommand \BibitemOpen [0]{}%
\providecommand \bibitemStop [0]{}%
\providecommand \bibitemNoStop [0]{.\EOS\space}%
\providecommand \EOS [0]{\spacefactor3000\relax}%
\providecommand \BibitemShut  [1]{\csname bibitem#1\endcsname}%
\let\auto@bib@innerbib\@empty
\bibitem [{\citenamefont {Einstein}\ \emph {et~al.}(1935)\citenamefont
  {Einstein}, \citenamefont {Podolsky},\ and\ \citenamefont {Rosen}}]{epr}%
  \BibitemOpen
  \bibfield  {author} {\bibinfo {author} {\bibfnamefont {A.}~\bibnamefont
  {Einstein}}, \bibinfo {author} {\bibfnamefont {B.}~\bibnamefont {Podolsky}},\
  and\ \bibinfo {author} {\bibfnamefont {N.}~\bibnamefont {Rosen}},\ }\bibfield
   {title} {\bibinfo {title} {Can quantum-mechanical description of physical
  reality be considered complete?},\ }\href
  {https://doi.org/10.1103/PhysRev.47.777} {\bibfield  {journal} {\bibinfo
  {journal} {Phys. Rev.}\ }\textbf {\bibinfo {volume} {47}},\ \bibinfo {pages}
  {777} (\bibinfo {year} {1935})}\BibitemShut {NoStop}%
\bibitem [{\citenamefont {Wiseman}\ \emph {et~al.}(2007)\citenamefont
  {Wiseman}, \citenamefont {Jones},\ and\ \citenamefont
  {Doherty}}]{wisemanSteering2007}%
  \BibitemOpen
  \bibfield  {author} {\bibinfo {author} {\bibfnamefont {H.~M.}\ \bibnamefont
  {Wiseman}}, \bibinfo {author} {\bibfnamefont {S.~J.}\ \bibnamefont {Jones}},\
  and\ \bibinfo {author} {\bibfnamefont {A.~C.}\ \bibnamefont {Doherty}},\
  }\bibfield  {title} {\bibinfo {title} {Steering, entanglement, nonlocality,
  and the {E}instein-{P}odolsky-{R}osen paradox},\ }\href
  {https://doi.org/10.1103/PhysRevLett.98.140402} {\bibfield  {journal}
  {\bibinfo  {journal} {Phys. Rev. Lett.}\ }\textbf {\bibinfo {volume} {98}},\
  \bibinfo {pages} {140402} (\bibinfo {year} {2007})}\BibitemShut {NoStop}%
\bibitem [{\citenamefont {Jones}\ \emph {et~al.}(2007)\citenamefont {Jones},
  \citenamefont {Wiseman},\ and\ \citenamefont {Doherty}}]{jones2007}%
  \BibitemOpen
  \bibfield  {author} {\bibinfo {author} {\bibfnamefont {S.}~\bibnamefont
  {Jones}}, \bibinfo {author} {\bibfnamefont {H.}~\bibnamefont {Wiseman}},\
  and\ \bibinfo {author} {\bibfnamefont {A.}~\bibnamefont {Doherty}},\
  }\bibfield  {title} {\bibinfo {title} {Entanglement,
  {{Einstein}}-{{Podolsky}}-{{Rosen}} correlations, {{Bell}} nonlocality, and
  steering},\ }\href {https://doi.org/10.1103/PhysRevA.76.052116} {\bibfield
  {journal} {\bibinfo  {journal} {Phys. Rev. A}\ }\textbf {\bibinfo {volume}
  {76}},\ \bibinfo {pages} {052116} (\bibinfo {year} {2007})}\BibitemShut
  {NoStop}%
\bibitem [{\citenamefont {Saunders}\ \emph {et~al.}(2010)\citenamefont
  {Saunders}, \citenamefont {Jones}, \citenamefont {Wiseman},\ and\
  \citenamefont {Pryde}}]{saunders2010}%
  \BibitemOpen
  \bibfield  {author} {\bibinfo {author} {\bibfnamefont {D.~J.}\ \bibnamefont
  {Saunders}}, \bibinfo {author} {\bibfnamefont {S.~J.}\ \bibnamefont {Jones}},
  \bibinfo {author} {\bibfnamefont {H.~M.}\ \bibnamefont {Wiseman}},\ and\
  \bibinfo {author} {\bibfnamefont {G.~J.}\ \bibnamefont {Pryde}},\ }\bibfield
  {title} {\bibinfo {title} {Experimental {{EPR}}-steering using {{Bell}}-local
  states},\ }\href {https://doi.org/10.1038/nphys1766} {\bibfield  {journal}
  {\bibinfo  {journal} {Nat. Phys.}\ }\textbf {\bibinfo {volume} {6}},\
  \bibinfo {pages} {845} (\bibinfo {year} {2010})}\BibitemShut {NoStop}%
\bibitem [{\citenamefont {Uola}\ \emph {et~al.}(2020)\citenamefont {Uola},
  \citenamefont {Costa}, \citenamefont {Nguyen},\ and\ \citenamefont
  {G\"uhne}}]{Uola2020}%
  \BibitemOpen
  \bibfield  {author} {\bibinfo {author} {\bibfnamefont {R.}~\bibnamefont
  {Uola}}, \bibinfo {author} {\bibfnamefont {A.~C.~S.}\ \bibnamefont {Costa}},
  \bibinfo {author} {\bibfnamefont {H.~C.}\ \bibnamefont {Nguyen}},\ and\
  \bibinfo {author} {\bibfnamefont {O.}~\bibnamefont {G\"uhne}},\ }\bibfield
  {title} {\bibinfo {title} {Quantum steering},\ }\href
  {https://doi.org/10.1103/RevModPhys.92.015001} {\bibfield  {journal}
  {\bibinfo  {journal} {Rev. Mod. Phys.}\ }\textbf {\bibinfo {volume} {92}},\
  \bibinfo {pages} {015001} (\bibinfo {year} {2020})}\BibitemShut {NoStop}%
\bibitem [{\citenamefont {Horodecki}\ \emph {et~al.}(2009)\citenamefont
  {Horodecki}, \citenamefont {Horodecki}, \citenamefont {Horodecki},\ and\
  \citenamefont {Horodecki}}]{horodecki2009}%
  \BibitemOpen
  \bibfield  {author} {\bibinfo {author} {\bibfnamefont {R.}~\bibnamefont
  {Horodecki}}, \bibinfo {author} {\bibfnamefont {P.}~\bibnamefont
  {Horodecki}}, \bibinfo {author} {\bibfnamefont {M.}~\bibnamefont
  {Horodecki}},\ and\ \bibinfo {author} {\bibfnamefont {K.}~\bibnamefont
  {Horodecki}},\ }\bibfield  {title} {\bibinfo {title} {Quantum entanglement},\
  }\href {https://doi.org/10.1103/RevModPhys.81.865} {\bibfield  {journal}
  {\bibinfo  {journal} {Rev. Mod. Phys.}\ }\textbf {\bibinfo {volume} {81}},\
  \bibinfo {pages} {865} (\bibinfo {year} {2009})}\BibitemShut {NoStop}%
\bibitem [{\citenamefont {Brunner}\ \emph {et~al.}(2014)\citenamefont
  {Brunner}, \citenamefont {Cavalcanti}, \citenamefont {Pironio}, \citenamefont
  {Scarani},\ and\ \citenamefont {Wehner}}]{brunner2014}%
  \BibitemOpen
  \bibfield  {author} {\bibinfo {author} {\bibfnamefont {N.}~\bibnamefont
  {Brunner}}, \bibinfo {author} {\bibfnamefont {D.}~\bibnamefont {Cavalcanti}},
  \bibinfo {author} {\bibfnamefont {S.}~\bibnamefont {Pironio}}, \bibinfo
  {author} {\bibfnamefont {V.}~\bibnamefont {Scarani}},\ and\ \bibinfo {author}
  {\bibfnamefont {S.}~\bibnamefont {Wehner}},\ }\bibfield  {title} {\bibinfo
  {title} {Bell nonlocality},\ }\href
  {https://doi.org/10.1103/RevModPhys.86.419} {\bibfield  {journal} {\bibinfo
  {journal} {Rev. Mod. Phys.}\ }\textbf {\bibinfo {volume} {86}},\ \bibinfo
  {pages} {419} (\bibinfo {year} {2014})}\BibitemShut {NoStop}%
\bibitem [{\citenamefont {Branciard}\ \emph {et~al.}(2012)\citenamefont
  {Branciard}, \citenamefont {Cavalcanti}, \citenamefont {Walborn},
  \citenamefont {Scarani},\ and\ \citenamefont {Wiseman}}]{branciard2012}%
  \BibitemOpen
  \bibfield  {author} {\bibinfo {author} {\bibfnamefont {C.}~\bibnamefont
  {Branciard}}, \bibinfo {author} {\bibfnamefont {E.~G.}\ \bibnamefont
  {Cavalcanti}}, \bibinfo {author} {\bibfnamefont {S.~P.}\ \bibnamefont
  {Walborn}}, \bibinfo {author} {\bibfnamefont {V.}~\bibnamefont {Scarani}},\
  and\ \bibinfo {author} {\bibfnamefont {H.~M.}\ \bibnamefont {Wiseman}},\
  }\bibfield  {title} {\bibinfo {title} {One-sided device-independent quantum
  key distribution: {{Security}}, feasibility, and the connection with
  steering},\ }\href {https://doi.org/10.1103/PhysRevA.85.010301} {\bibfield
  {journal} {\bibinfo  {journal} {Phys. Rev. A}\ }\textbf {\bibinfo {volume}
  {85}},\ \bibinfo {pages} {010301} (\bibinfo {year} {2012})}\BibitemShut
  {NoStop}%
\bibitem [{\citenamefont {Skrzypczyk}\ and\ \citenamefont
  {Cavalcanti}(2018)}]{skrzypczyk2018}%
  \BibitemOpen
  \bibfield  {author} {\bibinfo {author} {\bibfnamefont {P.}~\bibnamefont
  {Skrzypczyk}}\ and\ \bibinfo {author} {\bibfnamefont {D.}~\bibnamefont
  {Cavalcanti}},\ }\bibfield  {title} {\bibinfo {title} {Maximal {{Randomness
  Generation}} from {{Steering Inequality Violations Using Qudits}}},\ }\href
  {https://doi.org/10.1103/PhysRevLett.120.260401} {\bibfield  {journal}
  {\bibinfo  {journal} {Phys. Rev. Lett.}\ }\textbf {\bibinfo {volume} {120}},\
  \bibinfo {pages} {260401} (\bibinfo {year} {2018})}\BibitemShut {NoStop}%
\bibitem [{\citenamefont {Passaro}\ \emph {et~al.}(2015)\citenamefont
  {Passaro}, \citenamefont {Cavalcanti}, \citenamefont {Skrzypczyk},\ and\
  \citenamefont {Ac{\'i}n}}]{passaro2015Optimal}%
  \BibitemOpen
  \bibfield  {author} {\bibinfo {author} {\bibfnamefont {E.}~\bibnamefont
  {Passaro}}, \bibinfo {author} {\bibfnamefont {D.}~\bibnamefont {Cavalcanti}},
  \bibinfo {author} {\bibfnamefont {P.}~\bibnamefont {Skrzypczyk}},\ and\
  \bibinfo {author} {\bibfnamefont {A.}~\bibnamefont {Ac{\'i}n}},\ }\bibfield
  {title} {\bibinfo {title} {Optimal randomness certification in the quantum
  steering and prepare-and-measure scenarios},\ }\href
  {https://doi.org/10.1088/1367-2630/17/11/113010} {\bibfield  {journal}
  {\bibinfo  {journal} {New J. Phys.}\ }\textbf {\bibinfo {volume} {17}},\
  \bibinfo {pages} {113010} (\bibinfo {year} {2015})}\BibitemShut {NoStop}%
\bibitem [{\citenamefont {Piani}\ and\ \citenamefont
  {Watrous}(2015)}]{piani2015}%
  \BibitemOpen
  \bibfield  {author} {\bibinfo {author} {\bibfnamefont {M.}~\bibnamefont
  {Piani}}\ and\ \bibinfo {author} {\bibfnamefont {J.}~\bibnamefont
  {Watrous}},\ }\bibfield  {title} {\bibinfo {title} {Necessary and
  {{Sufficient Quantum Information Characterization}} of
  {{Einstein}}-{{Podolsky}}-{{Rosen Steering}}},\ }\href
  {https://doi.org/10.1103/PhysRevLett.114.060404} {\bibfield  {journal}
  {\bibinfo  {journal} {Phys. Rev. Lett.}\ }\textbf {\bibinfo {volume} {114}},\
  \bibinfo {pages} {060404} (\bibinfo {year} {2015})}\BibitemShut {NoStop}%
\bibitem [{\citenamefont {Reid}(2013)}]{reid2013}%
  \BibitemOpen
  \bibfield  {author} {\bibinfo {author} {\bibfnamefont {M.}~\bibnamefont
  {Reid}},\ }\bibfield  {title} {\bibinfo {title} {Signifying quantum
  benchmarks for qubit teleportation and secure quantum communication using
  {{Einstein}}-{{Podolsky}}-{{Rosen}} steering inequalities},\ }\href
  {https://doi.org/10.1103/PhysRevA.88.062338} {\bibfield  {journal} {\bibinfo
  {journal} {Phys. Rev. A}\ }\textbf {\bibinfo {volume} {88}},\ \bibinfo
  {pages} {062338} (\bibinfo {year} {2013})}\BibitemShut {NoStop}%
\bibitem [{\citenamefont {He}\ \emph {et~al.}(2015)\citenamefont {He},
  \citenamefont {{Rosales-Z{\'a}rate}}, \citenamefont {Adesso},\ and\
  \citenamefont {Reid}}]{he2015Secure}%
  \BibitemOpen
  \bibfield  {author} {\bibinfo {author} {\bibfnamefont {Q.}~\bibnamefont
  {He}}, \bibinfo {author} {\bibfnamefont {L.}~\bibnamefont
  {{Rosales-Z{\'a}rate}}}, \bibinfo {author} {\bibfnamefont {G.}~\bibnamefont
  {Adesso}},\ and\ \bibinfo {author} {\bibfnamefont {M.~D.}\ \bibnamefont
  {Reid}},\ }\bibfield  {title} {\bibinfo {title} {Secure {{Continuous Variable
  Teleportation}} and {{Einstein}}-{{Podolsky}}-{{Rosen Steering}}},\ }\href
  {https://doi.org/10.1103/PhysRevLett.115.180502} {\bibfield  {journal}
  {\bibinfo  {journal} {Phys. Rev. Lett.}\ }\textbf {\bibinfo {volume} {115}},\
  \bibinfo {pages} {180502} (\bibinfo {year} {2015})}\BibitemShut {NoStop}%
\bibitem [{\citenamefont {Cavalcanti}\ \emph {et~al.}(2013)\citenamefont
  {Cavalcanti}, \citenamefont {Hall},\ and\ \citenamefont
  {Wiseman}}]{cavalcantiE2013}%
  \BibitemOpen
  \bibfield  {author} {\bibinfo {author} {\bibfnamefont {E.~G.}\ \bibnamefont
  {Cavalcanti}}, \bibinfo {author} {\bibfnamefont {M.~J.~W.}\ \bibnamefont
  {Hall}},\ and\ \bibinfo {author} {\bibfnamefont {H.~M.}\ \bibnamefont
  {Wiseman}},\ }\bibfield  {title} {\bibinfo {title} {Entanglement verification
  and steering when {{Alice}} and {{Bob}} cannot be trusted},\ }\href
  {https://doi.org/10.1103/PhysRevA.87.032306} {\bibfield  {journal} {\bibinfo
  {journal} {Phys. Rev. A}\ }\textbf {\bibinfo {volume} {87}},\ \bibinfo
  {pages} {032306} (\bibinfo {year} {2013})}\BibitemShut {NoStop}%
\bibitem [{\citenamefont {Chen}\ \emph {et~al.}(2013)\citenamefont {Chen},
  \citenamefont {Ye}, \citenamefont {Wu}, \citenamefont {Su}, \citenamefont
  {Cabello}, \citenamefont {Kwek},\ and\ \citenamefont {Oh}}]{chenAll2013}%
  \BibitemOpen
  \bibfield  {author} {\bibinfo {author} {\bibfnamefont {J.-L.}\ \bibnamefont
  {Chen}}, \bibinfo {author} {\bibfnamefont {X.-J.}\ \bibnamefont {Ye}},
  \bibinfo {author} {\bibfnamefont {C.}~\bibnamefont {Wu}}, \bibinfo {author}
  {\bibfnamefont {H.-Y.}\ \bibnamefont {Su}}, \bibinfo {author} {\bibfnamefont
  {A.}~\bibnamefont {Cabello}}, \bibinfo {author} {\bibfnamefont {L.~C.}\
  \bibnamefont {Kwek}},\ and\ \bibinfo {author} {\bibfnamefont {C.~H.}\
  \bibnamefont {Oh}},\ }\bibfield  {title} {\bibinfo {title}
  {All-versus-nothing proof of {{Einstein}}-{{Podolsky}}-{{Rosen}} steering},\
  }\href {https://doi.org/10.1038/srep02143} {\bibfield  {journal} {\bibinfo
  {journal} {Sci. Rep.}\ }\textbf {\bibinfo {volume} {3}},\ \bibinfo {pages}
  {02143} (\bibinfo {year} {2013})}\BibitemShut {NoStop}%
\bibitem [{\citenamefont {Evans}\ and\ \citenamefont
  {Wiseman}(2014)}]{evans2014}%
  \BibitemOpen
  \bibfield  {author} {\bibinfo {author} {\bibfnamefont {D.~A.}\ \bibnamefont
  {Evans}}\ and\ \bibinfo {author} {\bibfnamefont {H.~M.}\ \bibnamefont
  {Wiseman}},\ }\bibfield  {title} {\bibinfo {title} {Optimal measurements for
  tests of {{Einstein}}-{{Podolsky}}-{{Rosen}} steering with no detection
  loophole using two-qubit {{Werner}} states},\ }\href
  {https://doi.org/10.1103/PhysRevA.90.012114} {\bibfield  {journal} {\bibinfo
  {journal} {Phys. Rev. A}\ }\textbf {\bibinfo {volume} {90}},\ \bibinfo
  {pages} {012114} (\bibinfo {year} {2014})}\BibitemShut {NoStop}%
\bibitem [{\citenamefont {Cavalcanti}\ \emph {et~al.}(2015)\citenamefont
  {Cavalcanti}, \citenamefont {Skrzypczyk}, \citenamefont {Aguilar},
  \citenamefont {Nery}, \citenamefont {Ribeiro},\ and\ \citenamefont
  {Walborn}}]{cavalcantiD2015}%
  \BibitemOpen
  \bibfield  {author} {\bibinfo {author} {\bibfnamefont {D.}~\bibnamefont
  {Cavalcanti}}, \bibinfo {author} {\bibfnamefont {P.}~\bibnamefont
  {Skrzypczyk}}, \bibinfo {author} {\bibfnamefont {G.~H.}\ \bibnamefont
  {Aguilar}}, \bibinfo {author} {\bibfnamefont {R.~V.}\ \bibnamefont {Nery}},
  \bibinfo {author} {\bibfnamefont {P.~H.~S.}\ \bibnamefont {Ribeiro}},\ and\
  \bibinfo {author} {\bibfnamefont {S.~P.}\ \bibnamefont {Walborn}},\
  }\bibfield  {title} {\bibinfo {title} {Detection of entanglement in
  asymmetric quantum networks and multipartite quantum steering},\ }\href
  {https://doi.org/10.1038/ncomms8941} {\bibfield  {journal} {\bibinfo
  {journal} {Nat. Commun.}\ }\textbf {\bibinfo {volume} {6}},\ \bibinfo {pages}
  {7941} (\bibinfo {year} {2015})}\BibitemShut {NoStop}%
\bibitem [{\citenamefont {Baker}\ \emph {et~al.}(2018)\citenamefont {Baker},
  \citenamefont {Wollmann}, \citenamefont {Pryde},\ and\ \citenamefont
  {Wiseman}}]{baker2018}%
  \BibitemOpen
  \bibfield  {author} {\bibinfo {author} {\bibfnamefont {T.~J.}\ \bibnamefont
  {Baker}}, \bibinfo {author} {\bibfnamefont {S.}~\bibnamefont {Wollmann}},
  \bibinfo {author} {\bibfnamefont {G.~J.}\ \bibnamefont {Pryde}},\ and\
  \bibinfo {author} {\bibfnamefont {H.~M.}\ \bibnamefont {Wiseman}},\
  }\bibfield  {title} {\bibinfo {title} {Necessary condition for steerability
  of arbitrary two-qubit states with loss},\ }\href
  {https://doi.org/10.1088/2040-8986/aaaa3c} {\bibfield  {journal} {\bibinfo
  {journal} {J. Opt.}\ }\textbf {\bibinfo {volume} {20}},\ \bibinfo {pages}
  {034008} (\bibinfo {year} {2018})}\BibitemShut {NoStop}%
\bibitem [{\citenamefont {Bowles}\ \emph {et~al.}(2014)\citenamefont {Bowles},
  \citenamefont {V{\'e}rtesi}, \citenamefont {Quintino},\ and\ \citenamefont
  {Brunner}}]{bowlesOneway2014}%
  \BibitemOpen
  \bibfield  {author} {\bibinfo {author} {\bibfnamefont {J.}~\bibnamefont
  {Bowles}}, \bibinfo {author} {\bibfnamefont {T.}~\bibnamefont {V{\'e}rtesi}},
  \bibinfo {author} {\bibfnamefont {M.~T.}\ \bibnamefont {Quintino}},\ and\
  \bibinfo {author} {\bibfnamefont {N.}~\bibnamefont {Brunner}},\ }\bibfield
  {title} {\bibinfo {title} {One-way {{Einstein}}-{{Podolsky}}-{{Rosen
  Steering}}},\ }\href {https://doi.org/10.1103/PhysRevLett.112.200402}
  {\bibfield  {journal} {\bibinfo  {journal} {Phys. Rev. Lett.}\ }\textbf
  {\bibinfo {volume} {112}},\ \bibinfo {pages} {200402} (\bibinfo {year}
  {2014})}\BibitemShut {NoStop}%
\bibitem [{\citenamefont {Bowles}\ \emph {et~al.}(2016)\citenamefont {Bowles},
  \citenamefont {Hirsch}, \citenamefont {Quintino},\ and\ \citenamefont
  {Brunner}}]{bowlesSufficient2016}%
  \BibitemOpen
  \bibfield  {author} {\bibinfo {author} {\bibfnamefont {J.}~\bibnamefont
  {Bowles}}, \bibinfo {author} {\bibfnamefont {F.}~\bibnamefont {Hirsch}},
  \bibinfo {author} {\bibfnamefont {M.~T.}\ \bibnamefont {Quintino}},\ and\
  \bibinfo {author} {\bibfnamefont {N.}~\bibnamefont {Brunner}},\ }\bibfield
  {title} {\bibinfo {title} {Sufficient criterion for guaranteeing that a
  two-qubit state is unsteerable},\ }\href
  {https://doi.org/10.1103/PhysRevA.93.022121} {\bibfield  {journal} {\bibinfo
  {journal} {Phys. Rev. A}\ }\textbf {\bibinfo {volume} {93}},\ \bibinfo
  {pages} {022121} (\bibinfo {year} {2016})}\BibitemShut {NoStop}%
\bibitem [{\citenamefont {Zhu}\ \emph {et~al.}(2016)\citenamefont {Zhu},
  \citenamefont {Hayashi},\ and\ \citenamefont {Chen}}]{zhu2016}%
  \BibitemOpen
  \bibfield  {author} {\bibinfo {author} {\bibfnamefont {H.}~\bibnamefont
  {Zhu}}, \bibinfo {author} {\bibfnamefont {M.}~\bibnamefont {Hayashi}},\ and\
  \bibinfo {author} {\bibfnamefont {L.}~\bibnamefont {Chen}},\ }\bibfield
  {title} {\bibinfo {title} {Universal {{Steering Criteria}}},\ }\href
  {https://doi.org/10.1103/PhysRevLett.116.070403} {\bibfield  {journal}
  {\bibinfo  {journal} {Phys. Rev. Lett.}\ }\textbf {\bibinfo {volume} {116}},\
  \bibinfo {pages} {070403} (\bibinfo {year} {2016})}\BibitemShut {NoStop}%
\bibitem [{\citenamefont {Skrzypczyk}\ \emph {et~al.}(2014)\citenamefont
  {Skrzypczyk}, \citenamefont {Navascu{\'e}s},\ and\ \citenamefont
  {Cavalcanti}}]{skr2014}%
  \BibitemOpen
  \bibfield  {author} {\bibinfo {author} {\bibfnamefont {P.}~\bibnamefont
  {Skrzypczyk}}, \bibinfo {author} {\bibfnamefont {M.}~\bibnamefont
  {Navascu{\'e}s}},\ and\ \bibinfo {author} {\bibfnamefont {D.}~\bibnamefont
  {Cavalcanti}},\ }\bibfield  {title} {\bibinfo {title} {Quantifying
  {{Einstein}}-{{Podolsky}}-{{Rosen Steering}}},\ }\href
  {https://doi.org/10.1103/PhysRevLett.112.180404} {\bibfield  {journal}
  {\bibinfo  {journal} {Phys. Rev. Lett.}\ }\textbf {\bibinfo {volume} {112}},\
  \bibinfo {pages} {180404} (\bibinfo {year} {2014})}\BibitemShut {NoStop}%
\bibitem [{\citenamefont {Quintino}\ \emph {et~al.}(2015)\citenamefont
  {Quintino}, \citenamefont {V{\'e}rtesi}, \citenamefont {Cavalcanti},
  \citenamefont {Augusiak}, \citenamefont {Demianowicz}, \citenamefont
  {Ac{\'i}n},\ and\ \citenamefont {Brunner}}]{quintino2015}%
  \BibitemOpen
  \bibfield  {author} {\bibinfo {author} {\bibfnamefont {M.~T.}\ \bibnamefont
  {Quintino}}, \bibinfo {author} {\bibfnamefont {T.}~\bibnamefont
  {V{\'e}rtesi}}, \bibinfo {author} {\bibfnamefont {D.}~\bibnamefont
  {Cavalcanti}}, \bibinfo {author} {\bibfnamefont {R.}~\bibnamefont
  {Augusiak}}, \bibinfo {author} {\bibfnamefont {M.}~\bibnamefont
  {Demianowicz}}, \bibinfo {author} {\bibfnamefont {A.}~\bibnamefont
  {Ac{\'i}n}},\ and\ \bibinfo {author} {\bibfnamefont {N.}~\bibnamefont
  {Brunner}},\ }\bibfield  {title} {\bibinfo {title} {Inequivalence of
  entanglement, steering, and {{Bell}} nonlocality for general measurements},\
  }\href {https://doi.org/10.1103/PhysRevA.92.032107} {\bibfield  {journal}
  {\bibinfo  {journal} {Phys. Rev. A}\ }\textbf {\bibinfo {volume} {92}},\
  \bibinfo {pages} {032107} (\bibinfo {year} {2015})}\BibitemShut {NoStop}%
\bibitem [{\citenamefont {Nguyen}\ \emph {et~al.}(2019)\citenamefont {Nguyen},
  \citenamefont {Nguyen},\ and\ \citenamefont {G\"uhne}}]{nguyengeometry2019}%
  \BibitemOpen
  \bibfield  {author} {\bibinfo {author} {\bibfnamefont {H.~C.}\ \bibnamefont
  {Nguyen}}, \bibinfo {author} {\bibfnamefont {H.-V.}\ \bibnamefont {Nguyen}},\
  and\ \bibinfo {author} {\bibfnamefont {O.}~\bibnamefont {G\"uhne}},\
  }\bibfield  {title} {\bibinfo {title} {Geometry of
  {E}instein-{P}odolsky-{R}osen correlations},\ }\href
  {https://doi.org/10.1103/PhysRevLett.122.240401} {\bibfield  {journal}
  {\bibinfo  {journal} {Phys. Rev. Lett.}\ }\textbf {\bibinfo {volume} {122}},\
  \bibinfo {pages} {240401} (\bibinfo {year} {2019})}\BibitemShut {NoStop}%
\bibitem [{\citenamefont {Baker}\ and\ \citenamefont
  {Wiseman}(2020)}]{bakerN2020}%
  \BibitemOpen
  \bibfield  {author} {\bibinfo {author} {\bibfnamefont {T.~J.}\ \bibnamefont
  {Baker}}\ and\ \bibinfo {author} {\bibfnamefont {H.~M.}\ \bibnamefont
  {Wiseman}},\ }\bibfield  {title} {\bibinfo {title} {Necessary conditions for
  steerability of two qubits from consideration of local operations},\ }\href
  {https://doi.org/10.1103/PhysRevA.101.022326} {\bibfield  {journal} {\bibinfo
   {journal} {Phys. Rev. A}\ }\textbf {\bibinfo {volume} {101}},\ \bibinfo
  {pages} {022326} (\bibinfo {year} {2020})}\BibitemShut {NoStop}%
\bibitem [{\citenamefont {Reid}(1989)}]{reid1989}%
  \BibitemOpen
  \bibfield  {author} {\bibinfo {author} {\bibfnamefont {M.}~\bibnamefont
  {Reid}},\ }\bibfield  {title} {\bibinfo {title} {Demonstration of the
  {{Einstein}}-{{Podolsky}}-{{Rosen}} paradox using nondegenerate parametric
  amplification},\ }\href {https://doi.org/10.1103/PhysRevA.40.913} {\bibfield
  {journal} {\bibinfo  {journal} {Phys. Rev. A}\ }\textbf {\bibinfo {volume}
  {40}},\ \bibinfo {pages} {913} (\bibinfo {year} {1989})}\BibitemShut
  {NoStop}%
\bibitem [{\citenamefont {Midgley}\ \emph {et~al.}(2010)\citenamefont
  {Midgley}, \citenamefont {Ferris},\ and\ \citenamefont
  {Olsen}}]{midgley2010}%
  \BibitemOpen
  \bibfield  {author} {\bibinfo {author} {\bibfnamefont {S.~L.~W.}\
  \bibnamefont {Midgley}}, \bibinfo {author} {\bibfnamefont {A.~J.}\
  \bibnamefont {Ferris}},\ and\ \bibinfo {author} {\bibfnamefont {M.~K.}\
  \bibnamefont {Olsen}},\ }\bibfield  {title} {\bibinfo {title} {Asymmetric
  {{Gaussian}} steering: {{When Alice}} and {{Bob}} disagree},\ }\href
  {https://doi.org/10.1103/PhysRevA.81.022101} {\bibfield  {journal} {\bibinfo
  {journal} {Phys. Rev. A}\ }\textbf {\bibinfo {volume} {81}},\ \bibinfo
  {pages} {022101} (\bibinfo {year} {2010})}\BibitemShut {NoStop}%
\bibitem [{\citenamefont {Wagner}\ \emph {et~al.}(2008)\citenamefont {Wagner},
  \citenamefont {Janousek}, \citenamefont {Delaubert}, \citenamefont {Zou},
  \citenamefont {Harb}, \citenamefont {Treps}, \citenamefont {Morizur},
  \citenamefont {Lam},\ and\ \citenamefont {Bachor}}]{wagner2008}%
  \BibitemOpen
  \bibfield  {author} {\bibinfo {author} {\bibfnamefont {K.}~\bibnamefont
  {Wagner}}, \bibinfo {author} {\bibfnamefont {J.}~\bibnamefont {Janousek}},
  \bibinfo {author} {\bibfnamefont {V.}~\bibnamefont {Delaubert}}, \bibinfo
  {author} {\bibfnamefont {H.}~\bibnamefont {Zou}}, \bibinfo {author}
  {\bibfnamefont {C.}~\bibnamefont {Harb}}, \bibinfo {author} {\bibfnamefont
  {N.}~\bibnamefont {Treps}}, \bibinfo {author} {\bibfnamefont {J.~F.}\
  \bibnamefont {Morizur}}, \bibinfo {author} {\bibfnamefont {P.~K.}\
  \bibnamefont {Lam}},\ and\ \bibinfo {author} {\bibfnamefont {H.~A.}\
  \bibnamefont {Bachor}},\ }\bibfield  {title} {\bibinfo {title} {Entangling
  the spatial properties of laser beams},\ }\href
  {https://doi.org/10.1126/science.1159663} {\bibfield  {journal} {\bibinfo
  {journal} {Science}\ }\textbf {\bibinfo {volume} {321}},\ \bibinfo {pages}
  {541} (\bibinfo {year} {2008})}\BibitemShut {NoStop}%
\bibitem [{\citenamefont {H{\"a}ndchen}\ \emph {et~al.}(2012)\citenamefont
  {H{\"a}ndchen}, \citenamefont {Eberle}, \citenamefont {Steinlechner},
  \citenamefont {Samblowski}, \citenamefont {Franz}, \citenamefont {Werner},\
  and\ \citenamefont {Schnabel}}]{handchen2012}%
  \BibitemOpen
  \bibfield  {author} {\bibinfo {author} {\bibfnamefont {V.}~\bibnamefont
  {H{\"a}ndchen}}, \bibinfo {author} {\bibfnamefont {T.}~\bibnamefont
  {Eberle}}, \bibinfo {author} {\bibfnamefont {S.}~\bibnamefont
  {Steinlechner}}, \bibinfo {author} {\bibfnamefont {A.}~\bibnamefont
  {Samblowski}}, \bibinfo {author} {\bibfnamefont {T.}~\bibnamefont {Franz}},
  \bibinfo {author} {\bibfnamefont {R.~F.}\ \bibnamefont {Werner}},\ and\
  \bibinfo {author} {\bibfnamefont {R.}~\bibnamefont {Schnabel}},\ }\bibfield
  {title} {\bibinfo {title} {Observation of one-way
  {{Einstein}}\textendash{{Podolsky}}\textendash{{Rosen}} steering},\ }\href
  {https://doi.org/10.1038/nphoton.2012.202} {\bibfield  {journal} {\bibinfo
  {journal} {Nat. Photon.}\ }\textbf {\bibinfo {volume} {6}},\ \bibinfo {pages}
  {596} (\bibinfo {year} {2012})}\BibitemShut {NoStop}%
\bibitem [{\citenamefont {Chen}\ \emph {et~al.}(2002)\citenamefont {Chen},
  \citenamefont {Pan}, \citenamefont {Hou},\ and\ \citenamefont
  {Zhang}}]{chen2002}%
  \BibitemOpen
  \bibfield  {author} {\bibinfo {author} {\bibfnamefont {Z.-B.}\ \bibnamefont
  {Chen}}, \bibinfo {author} {\bibfnamefont {J.-W.}\ \bibnamefont {Pan}},
  \bibinfo {author} {\bibfnamefont {G.}~\bibnamefont {Hou}},\ and\ \bibinfo
  {author} {\bibfnamefont {Y.-D.}\ \bibnamefont {Zhang}},\ }\bibfield  {title}
  {\bibinfo {title} {Maximal violation of {B}ell's inequalities for continuous
  variable systems},\ }\href {https://doi.org/10.1103/PhysRevLett.88.040406}
  {\bibfield  {journal} {\bibinfo  {journal} {Phys. Rev. Lett.}\ }\textbf
  {\bibinfo {volume} {88}},\ \bibinfo {pages} {040406} (\bibinfo {year}
  {2002})}\BibitemShut {NoStop}%
\bibitem [{\citenamefont {Wollmann}\ \emph {et~al.}(2016)\citenamefont
  {Wollmann}, \citenamefont {Walk}, \citenamefont {Bennet}, \citenamefont
  {Wiseman},\ and\ \citenamefont {Pryde}}]{Wollmann2016}%
  \BibitemOpen
  \bibfield  {author} {\bibinfo {author} {\bibfnamefont {S.}~\bibnamefont
  {Wollmann}}, \bibinfo {author} {\bibfnamefont {N.}~\bibnamefont {Walk}},
  \bibinfo {author} {\bibfnamefont {A.~J.}\ \bibnamefont {Bennet}}, \bibinfo
  {author} {\bibfnamefont {H.~M.}\ \bibnamefont {Wiseman}},\ and\ \bibinfo
  {author} {\bibfnamefont {G.~J.}\ \bibnamefont {Pryde}},\ }\bibfield  {title}
  {\bibinfo {title} {Observation of genuine one-way
  {E}instein-{P}odolsky-{R}osen steering},\ }\href
  {https://doi.org/10.1103/PhysRevLett.116.160403} {\bibfield  {journal}
  {\bibinfo  {journal} {Phys. Rev. Lett.}\ }\textbf {\bibinfo {volume} {116}},\
  \bibinfo {pages} {160403} (\bibinfo {year} {2016})}\BibitemShut {NoStop}%
\bibitem [{\citenamefont {Sun}\ \emph {et~al.}(2016)\citenamefont {Sun},
  \citenamefont {Ye}, \citenamefont {Xu}, \citenamefont {Xu}, \citenamefont
  {Tang}, \citenamefont {Wu}, \citenamefont {Chen}, \citenamefont {Li},\ and\
  \citenamefont {Guo}}]{sun2016}%
  \BibitemOpen
  \bibfield  {author} {\bibinfo {author} {\bibfnamefont {K.}~\bibnamefont
  {Sun}}, \bibinfo {author} {\bibfnamefont {X.-J.}\ \bibnamefont {Ye}},
  \bibinfo {author} {\bibfnamefont {J.-S.}\ \bibnamefont {Xu}}, \bibinfo
  {author} {\bibfnamefont {X.-Y.}\ \bibnamefont {Xu}}, \bibinfo {author}
  {\bibfnamefont {J.-S.}\ \bibnamefont {Tang}}, \bibinfo {author}
  {\bibfnamefont {Y.-C.}\ \bibnamefont {Wu}}, \bibinfo {author} {\bibfnamefont
  {J.-L.}\ \bibnamefont {Chen}}, \bibinfo {author} {\bibfnamefont {C.-F.}\
  \bibnamefont {Li}},\ and\ \bibinfo {author} {\bibfnamefont {G.-C.}\
  \bibnamefont {Guo}},\ }\bibfield  {title} {\bibinfo {title} {Experimental
  quantification of asymmetric {E}instein-{P}odolsky-{R}osen steering},\ }\href
  {https://doi.org/10.1103/PhysRevLett.116.160404} {\bibfield  {journal}
  {\bibinfo  {journal} {Phys. Rev. Lett.}\ }\textbf {\bibinfo {volume} {116}},\
  \bibinfo {pages} {160404} (\bibinfo {year} {2016})}\BibitemShut {NoStop}%
\bibitem [{\citenamefont {Xiao}\ \emph {et~al.}(2017)\citenamefont {Xiao},
  \citenamefont {Ye}, \citenamefont {Sun}, \citenamefont {Xu}, \citenamefont
  {Li},\ and\ \citenamefont {Guo}}]{xiao2017}%
  \BibitemOpen
  \bibfield  {author} {\bibinfo {author} {\bibfnamefont {Y.}~\bibnamefont
  {Xiao}}, \bibinfo {author} {\bibfnamefont {X.-J.}\ \bibnamefont {Ye}},
  \bibinfo {author} {\bibfnamefont {K.}~\bibnamefont {Sun}}, \bibinfo {author}
  {\bibfnamefont {J.-S.}\ \bibnamefont {Xu}}, \bibinfo {author} {\bibfnamefont
  {C.-F.}\ \bibnamefont {Li}},\ and\ \bibinfo {author} {\bibfnamefont {G.-C.}\
  \bibnamefont {Guo}},\ }\bibfield  {title} {\bibinfo {title} {Demonstration of
  multisetting one-way {E}instein-{P}odolsky-{R}osen steering in two-qubit
  systems},\ }\href {https://doi.org/10.1103/PhysRevLett.118.140404} {\bibfield
   {journal} {\bibinfo  {journal} {Phys. Rev. Lett.}\ }\textbf {\bibinfo
  {volume} {118}},\ \bibinfo {pages} {140404} (\bibinfo {year}
  {2017})}\BibitemShut {NoStop}%
\bibitem [{\citenamefont {Tischler}\ \emph {et~al.}(2018)\citenamefont
  {Tischler}, \citenamefont {Ghafari}, \citenamefont {Baker}, \citenamefont
  {Slussarenko}, \citenamefont {Patel}, \citenamefont {Weston}, \citenamefont
  {Wollmann}, \citenamefont {Shalm}, \citenamefont {Verma}, \citenamefont
  {Nam}, \citenamefont {Nguyen}, \citenamefont {Wiseman},\ and\ \citenamefont
  {Pryde}}]{Tischler2018a}%
  \BibitemOpen
  \bibfield  {author} {\bibinfo {author} {\bibfnamefont {N.}~\bibnamefont
  {Tischler}}, \bibinfo {author} {\bibfnamefont {F.}~\bibnamefont {Ghafari}},
  \bibinfo {author} {\bibfnamefont {T.~J.}\ \bibnamefont {Baker}}, \bibinfo
  {author} {\bibfnamefont {S.}~\bibnamefont {Slussarenko}}, \bibinfo {author}
  {\bibfnamefont {R.~B.}\ \bibnamefont {Patel}}, \bibinfo {author}
  {\bibfnamefont {M.~M.}\ \bibnamefont {Weston}}, \bibinfo {author}
  {\bibfnamefont {S.}~\bibnamefont {Wollmann}}, \bibinfo {author}
  {\bibfnamefont {L.~K.}\ \bibnamefont {Shalm}}, \bibinfo {author}
  {\bibfnamefont {V.~B.}\ \bibnamefont {Verma}}, \bibinfo {author}
  {\bibfnamefont {S.~W.}\ \bibnamefont {Nam}}, \bibinfo {author} {\bibfnamefont
  {H.~C.}\ \bibnamefont {Nguyen}}, \bibinfo {author} {\bibfnamefont {H.~M.}\
  \bibnamefont {Wiseman}},\ and\ \bibinfo {author} {\bibfnamefont {G.~J.}\
  \bibnamefont {Pryde}},\ }\bibfield  {title} {\bibinfo {title} {Conclusive
  experimental demonstration of one-way {E}instein-{P}odolsky-{R}osen
  steering},\ }\href {https://doi.org/10.1103/PhysRevLett.121.100401}
  {\bibfield  {journal} {\bibinfo  {journal} {Phys. Rev. Lett.}\ }\textbf
  {\bibinfo {volume} {121}},\ \bibinfo {pages} {100401} (\bibinfo {year}
  {2018})}\BibitemShut {NoStop}%
\bibitem [{\citenamefont {Zeng}\ \emph {et~al.}(2018)\citenamefont {Zeng},
  \citenamefont {Wang}, \citenamefont {Li},\ and\ \citenamefont
  {Zhang}}]{zeng2018}%
  \BibitemOpen
  \bibfield  {author} {\bibinfo {author} {\bibfnamefont {Q.}~\bibnamefont
  {Zeng}}, \bibinfo {author} {\bibfnamefont {B.}~\bibnamefont {Wang}}, \bibinfo
  {author} {\bibfnamefont {P.}~\bibnamefont {Li}},\ and\ \bibinfo {author}
  {\bibfnamefont {X.}~\bibnamefont {Zhang}},\ }\bibfield  {title} {\bibinfo
  {title} {Experimental high-dimensional {E}instein-{P}odolsky-{R}osen
  steering},\ }\href {https://doi.org/10.1103/PhysRevLett.120.030401}
  {\bibfield  {journal} {\bibinfo  {journal} {Phys. Rev. Lett.}\ }\textbf
  {\bibinfo {volume} {120}},\ \bibinfo {pages} {030401} (\bibinfo {year}
  {2018})}\BibitemShut {NoStop}%
\bibitem [{\citenamefont {Qu}\ \emph {et~al.}()\citenamefont {Qu},
  \citenamefont {Wang}, \citenamefont {An}, \citenamefont {Wang}, \citenamefont
  {Li}, \citenamefont {Gao}, \citenamefont {Li},\ and\ \citenamefont
  {Zhang}}]{qu2021}%
  \BibitemOpen
  \bibfield  {author} {\bibinfo {author} {\bibfnamefont {R.}~\bibnamefont
  {Qu}}, \bibinfo {author} {\bibfnamefont {Y.}~\bibnamefont {Wang}}, \bibinfo
  {author} {\bibfnamefont {M.}~\bibnamefont {An}}, \bibinfo {author}
  {\bibfnamefont {F.}~\bibnamefont {Wang}}, \bibinfo {author} {\bibfnamefont
  {H.}~\bibnamefont {Li}}, \bibinfo {author} {\bibfnamefont {H.}~\bibnamefont
  {Gao}}, \bibinfo {author} {\bibfnamefont {F.}~\bibnamefont {Li}},\ and\
  \bibinfo {author} {\bibfnamefont {P.}~\bibnamefont {Zhang}},\ }\href@noop {}
  {\bibinfo {title} {Experimental demonstration of high-dimensional quantum
  steering with $n$ measurement settings}},\ \Eprint
  {https://arxiv.org/abs/2101.04436} {arXiv:2101.04436} \BibitemShut {NoStop}%
\bibitem [{\citenamefont {Karimi}\ \emph {et~al.}(2009)\citenamefont {Karimi},
  \citenamefont {Piccirillo}, \citenamefont {Nagali}, \citenamefont
  {Marrucci},\ and\ \citenamefont {Santamato}}]{karimiEfficient2009}%
  \BibitemOpen
  \bibfield  {author} {\bibinfo {author} {\bibfnamefont {E.}~\bibnamefont
  {Karimi}}, \bibinfo {author} {\bibfnamefont {B.}~\bibnamefont {Piccirillo}},
  \bibinfo {author} {\bibfnamefont {E.}~\bibnamefont {Nagali}}, \bibinfo
  {author} {\bibfnamefont {L.}~\bibnamefont {Marrucci}},\ and\ \bibinfo
  {author} {\bibfnamefont {E.}~\bibnamefont {Santamato}},\ }\bibfield  {title}
  {\bibinfo {title} {Efficient generation and sorting of orbital angular
  momentum eigenmodes of light by thermally tuned q-plates},\ }\href
  {https://doi.org/10.1063/1.3154549} {\bibfield  {journal} {\bibinfo
  {journal} {Appl. Phys. Lett.}\ }\textbf {\bibinfo {volume} {94}},\ \bibinfo
  {pages} {231124} (\bibinfo {year} {2009})}\BibitemShut {NoStop}%
\bibitem [{\citenamefont {Karimi}\ \emph {et~al.}(2010)\citenamefont {Karimi},
  \citenamefont {Leach}, \citenamefont {Slussarenko}, \citenamefont
  {Piccirillo}, \citenamefont {Marrucci}, \citenamefont {Chen}, \citenamefont
  {She}, \citenamefont {{Franke-Arnold}}, \citenamefont {Padgett},\ and\
  \citenamefont {Santamato}}]{karimiSpin2010}%
  \BibitemOpen
  \bibfield  {author} {\bibinfo {author} {\bibfnamefont {E.}~\bibnamefont
  {Karimi}}, \bibinfo {author} {\bibfnamefont {J.}~\bibnamefont {Leach}},
  \bibinfo {author} {\bibfnamefont {S.}~\bibnamefont {Slussarenko}}, \bibinfo
  {author} {\bibfnamefont {B.}~\bibnamefont {Piccirillo}}, \bibinfo {author}
  {\bibfnamefont {L.}~\bibnamefont {Marrucci}}, \bibinfo {author}
  {\bibfnamefont {L.}~\bibnamefont {Chen}}, \bibinfo {author} {\bibfnamefont
  {W.}~\bibnamefont {She}}, \bibinfo {author} {\bibfnamefont {S.}~\bibnamefont
  {{Franke-Arnold}}}, \bibinfo {author} {\bibfnamefont {M.~J.}\ \bibnamefont
  {Padgett}},\ and\ \bibinfo {author} {\bibfnamefont {E.}~\bibnamefont
  {Santamato}},\ }\bibfield  {title} {\bibinfo {title} {Spin-orbit hybrid
  entanglement of photons and quantum contextuality},\ }\href
  {https://doi.org/10.1103/PhysRevA.82.022115} {\bibfield  {journal} {\bibinfo
  {journal} {Phys. Rev. A}\ }\textbf {\bibinfo {volume} {82}},\ \bibinfo
  {pages} {022115} (\bibinfo {year} {2010})}\BibitemShut {NoStop}%
\bibitem [{\citenamefont {Slussarenko}\ \emph {et~al.}(2011)\citenamefont
  {Slussarenko}, \citenamefont {Karimi}, \citenamefont {Piccirillo},
  \citenamefont {Marrucci},\ and\ \citenamefont {Santamato}}]{slussarenko2011}%
  \BibitemOpen
  \bibfield  {author} {\bibinfo {author} {\bibfnamefont {S.}~\bibnamefont
  {Slussarenko}}, \bibinfo {author} {\bibfnamefont {E.}~\bibnamefont {Karimi}},
  \bibinfo {author} {\bibfnamefont {B.}~\bibnamefont {Piccirillo}}, \bibinfo
  {author} {\bibfnamefont {L.}~\bibnamefont {Marrucci}},\ and\ \bibinfo
  {author} {\bibfnamefont {E.}~\bibnamefont {Santamato}},\ }\bibfield  {title}
  {\bibinfo {title} {Efficient generation and control of different-order
  orbital angular momentum states for communication links},\ }\href
  {https://doi.org/10.1364/JOSAA.28.000061} {\bibfield  {journal} {\bibinfo
  {journal} {J. Opt. Soc. Am. A}\ }\textbf {\bibinfo {volume} {28}},\ \bibinfo
  {pages} {61} (\bibinfo {year} {2011})}\BibitemShut {NoStop}%
\bibitem [{\citenamefont {Zeng}\ \emph {et~al.}(2016)\citenamefont {Zeng},
  \citenamefont {Li}, \citenamefont {Song},\ and\ \citenamefont
  {Zhang}}]{zeng2016}%
  \BibitemOpen
  \bibfield  {author} {\bibinfo {author} {\bibfnamefont {Q.}~\bibnamefont
  {Zeng}}, \bibinfo {author} {\bibfnamefont {T.}~\bibnamefont {Li}}, \bibinfo
  {author} {\bibfnamefont {X.}~\bibnamefont {Song}},\ and\ \bibinfo {author}
  {\bibfnamefont {X.}~\bibnamefont {Zhang}},\ }\bibfield  {title} {\bibinfo
  {title} {Realization of optimized quantum controlled-logic gate based on the
  orbital angular momentum of light},\ }\href
  {https://doi.org/10.1364/OE.24.008186} {\bibfield  {journal} {\bibinfo
  {journal} {Opt. Express}\ }\textbf {\bibinfo {volume} {24}},\ \bibinfo
  {pages} {8186} (\bibinfo {year} {2016})}\BibitemShut {NoStop}%
\bibitem [{\citenamefont {Herrera~Valencia}\ \emph {et~al.}(2020)\citenamefont
  {Herrera~Valencia}, \citenamefont {Srivastav}, \citenamefont {Pivoluska},
  \citenamefont {Huber}, \citenamefont {Friis}, \citenamefont {McCutcheon},\
  and\ \citenamefont {Malik}}]{pixel2020}%
  \BibitemOpen
  \bibfield  {author} {\bibinfo {author} {\bibfnamefont {N.}~\bibnamefont
  {Herrera~Valencia}}, \bibinfo {author} {\bibfnamefont {V.}~\bibnamefont
  {Srivastav}}, \bibinfo {author} {\bibfnamefont {M.}~\bibnamefont
  {Pivoluska}}, \bibinfo {author} {\bibfnamefont {M.}~\bibnamefont {Huber}},
  \bibinfo {author} {\bibfnamefont {N.}~\bibnamefont {Friis}}, \bibinfo
  {author} {\bibfnamefont {W.}~\bibnamefont {McCutcheon}},\ and\ \bibinfo
  {author} {\bibfnamefont {M.}~\bibnamefont {Malik}},\ }\bibfield  {title}
  {\bibinfo {title} {High-{D}imensional {P}ixel {E}ntanglement: {E}fficient
  {G}eneration and {C}ertification},\ }\href
  {https://doi.org/10.22331/q-2020-12-24-376} {\bibfield  {journal} {\bibinfo
  {journal} {{Quantum}}\ }\textbf {\bibinfo {volume} {4}},\ \bibinfo {pages}
  {376} (\bibinfo {year} {2020})}\BibitemShut {NoStop}%
\bibitem [{\citenamefont {Grillo}\ \emph {et~al.}(2017)\citenamefont {Grillo},
  \citenamefont {Harvey}, \citenamefont {Venturi}, \citenamefont {Pierce},
  \citenamefont {Balboni}, \citenamefont {Bouchard}, \citenamefont
  {Carlo~Gazzadi}, \citenamefont {Frabboni}, \citenamefont {Tavabi},
  \citenamefont {Li}, \citenamefont {{Dunin-Borkowski}}, \citenamefont {Boyd},
  \citenamefont {McMorran},\ and\ \citenamefont
  {Karimi}}]{grillo2017Observation}%
  \BibitemOpen
  \bibfield  {author} {\bibinfo {author} {\bibfnamefont {V.}~\bibnamefont
  {Grillo}}, \bibinfo {author} {\bibfnamefont {T.~R.}\ \bibnamefont {Harvey}},
  \bibinfo {author} {\bibfnamefont {F.}~\bibnamefont {Venturi}}, \bibinfo
  {author} {\bibfnamefont {J.~S.}\ \bibnamefont {Pierce}}, \bibinfo {author}
  {\bibfnamefont {R.}~\bibnamefont {Balboni}}, \bibinfo {author} {\bibfnamefont
  {F.}~\bibnamefont {Bouchard}}, \bibinfo {author} {\bibfnamefont
  {G.}~\bibnamefont {Carlo~Gazzadi}}, \bibinfo {author} {\bibfnamefont
  {S.}~\bibnamefont {Frabboni}}, \bibinfo {author} {\bibfnamefont {A.~H.}\
  \bibnamefont {Tavabi}}, \bibinfo {author} {\bibfnamefont {Z.-A.}\
  \bibnamefont {Li}}, \bibinfo {author} {\bibfnamefont {R.~E.}\ \bibnamefont
  {{Dunin-Borkowski}}}, \bibinfo {author} {\bibfnamefont {R.~W.}\ \bibnamefont
  {Boyd}}, \bibinfo {author} {\bibfnamefont {B.~J.}\ \bibnamefont {McMorran}},\
  and\ \bibinfo {author} {\bibfnamefont {E.}~\bibnamefont {Karimi}},\
  }\bibfield  {title} {\bibinfo {title} {Observation of nanoscale magnetic
  fields using twisted electron beams},\ }\href
  {https://doi.org/10.1038/s41467-017-00829-5} {\bibfield  {journal} {\bibinfo
  {journal} {Nat. Commun.}\ }\textbf {\bibinfo {volume} {8}},\ \bibinfo {pages}
  {689} (\bibinfo {year} {2017})}\BibitemShut {NoStop}%
\bibitem [{\citenamefont {Agnew}\ \emph {et~al.}(2011)\citenamefont {Agnew},
  \citenamefont {Leach}, \citenamefont {McLaren}, \citenamefont {Roux},\ and\
  \citenamefont {Boyd}}]{PhysRevA.84.062101}%
  \BibitemOpen
  \bibfield  {author} {\bibinfo {author} {\bibfnamefont {M.}~\bibnamefont
  {Agnew}}, \bibinfo {author} {\bibfnamefont {J.}~\bibnamefont {Leach}},
  \bibinfo {author} {\bibfnamefont {M.}~\bibnamefont {McLaren}}, \bibinfo
  {author} {\bibfnamefont {F.~S.}\ \bibnamefont {Roux}},\ and\ \bibinfo
  {author} {\bibfnamefont {R.~W.}\ \bibnamefont {Boyd}},\ }\bibfield  {title}
  {\bibinfo {title} {Tomography of the quantum state of photons entangled in
  high dimensions},\ }\href {https://doi.org/10.1103/PhysRevA.84.062101}
  {\bibfield  {journal} {\bibinfo  {journal} {Phys. Rev. A}\ }\textbf {\bibinfo
  {volume} {84}},\ \bibinfo {pages} {062101} (\bibinfo {year}
  {2011})}\BibitemShut {NoStop}%
\end{thebibliography}


\providecommand{\noopsort}[1]{}\providecommand{\singleletter}[1]{#1}%
%


\end{document}